\documentclass[letterpaper]{article}
\usepackage{amsmath,amssymb}
\batchmode

\newtheorem{lem}{Lemma}
\newtheorem{thm}{Theorem}

\newtheorem{cor}{Corollary}

\newcommand{\De}{\Delta}
\newcommand{\de}{\delta}

\newcommand{\cB}{{\cal B}}
\newcommand{\cC}{{\cal C}}
\newcommand{\cS}{{\cal S}}
\newcommand{\cG}{{\cal G}}
\newcommand{\cL}{{\cal L}}
\newcommand{\cP}{{\cal P}}
\newcommand{\cZ}{{\cal Z}}

\begin{document}

\title{Infinite Volume Limit for Correlation functions\\ in the Dipole Gas}
\author{ 
Tuan Minh Le \footnote{e-mail addresses: tuanle@buffalo.edu { \hspace{.5cm}} or { \hspace{.5cm}} leminhtuan912@gmail.com}\\
Department  of Mathematics \\
SUNY at Buffalo \\
Buffalo, NY 14260 }
\maketitle

\begin{abstract}
We study a classical lattice  dipole gas  with low  activity  in dimension $d \geq 3$. We investigate long distance properties by a renormalization group analysis. We prove that various correlation functions have an infinite volume limit. We also get estimates on the decay of correlation functions.
\end{abstract} 

\section{Introduction}
\subsection{Overview}
In this paper we continue to study the classical dipole gas   on a unit lattice  ${\mathbb{Z}}^d$ with     $d \geq 3$. Each dipole is described by its position coordinate $x\in {\mathbb{Z}}^d$ and a unit polarization vector (moment) $p \in {\mathbb{S}}^{d-1}$.

 Let $\{e_1,\dots, e_d\}$ be the standard basic for ${\mathbb{Z}}^d$. For $\varphi: {\mathbb{Z}}^d \to {\mathbb{R}}$ and $ \mu \in \{1,\dots, d\}$  we define  $\partial_{\mu} \varphi$  as $\partial_{\mu} \varphi(x)  =  \varphi(x + e_{\mu}) -\varphi(x )$. 
Let   $e_{-\mu}  = - e_{\mu}$ with $\mu \in \{1,\dots,d\}$. Then the definition of $\partial_{\mu} \varphi$ can be used to define the forward or backward   lattice derivative along the unit  vector $e_{\mu}$ with $\mu \in \{\pm1,\dots,\pm d\}$.  We have that  $\partial_{\mu}$ and  $\partial_{-\mu}$   are adjoint to each other  and    $-\De  =  1/2  \sum_{ \pm \mu =1}^d \partial_{\mu}^*  \partial_{\mu} = 1/2  \sum_{\pm \mu=1}^d \partial_{-\mu}  \partial_{\mu}$.\footnote{We distinguish forward and backward derivatives to facilitate a symmetric  decomposition of  $V(\Lambda_N)$ (defined in (\ref{VLaN})) into  blocks.}

As in \cite{Dim09}, the  potential energy between  unit  dipoles $(x,p_1)$ and $(y,p_2)$ is  
\begin{equation}
(p_1 \cdot  \partial)   (p_2 \cdot  \partial)  C  (x-y)
\end{equation}  
where   $x,y  \in {\mathbb{Z}}^d$ are  positions,  $p_1, p_2 \in {\mathbb{S}}^{d-1}$ are    moments, $\partial = (\partial_1,\dots, \partial_d)$ and  $C(x-y)$ is the Coulomb potential on the unit lattice ${\mathbb{Z}}^d$, which  is   the kernel of the inverse Laplacian
\begin{equation}    
C(x,y)   =  (- \De)^{-1}  (x,y) =(2 \pi)^{-d}   \int_{[-\pi,\pi]^d}  \frac{e^{ip\cdot (x-y)}}{  2  \sum_{\mu=1}^d (1- \cos p_{\mu})} dp
\end{equation}    
And the potential energy of $n$ dipoles, including self energy,  has the form
\begin{equation}
\sum_{1 \leq k,j \leq n} (p_k \cdot \partial)(p_j \cdot \partial) C(x_k, x_j) 
\end{equation}  
Let      $\Lambda_N $ be a box  in  ${\mathbb{R}}^d$
\begin{equation}
   \Lambda_N  =  \left[ \frac{-L^N}{2},  \frac{L^N}{2}\right]^d
   \end{equation}
  where   $L  \geq 2^{d+3} +1 $ is a very  large, odd integer.    For    $\Lambda_N  \cap  {\mathbb{Z}}^d$, the classical statistical mechanics of a gas of such dipoles with    inverse temperature (for convenience)  $\beta =1$   and activity (fugacity)   $z>0$  is given  by  the  grand canonical   partition  function
\begin{equation} \label{grand}
\begin{split}
Z_N&= \sum_{n\geq 0} \frac{z^n}{n!} \prod_{i=1}^n  \sum_{x_i \in {\mathbb{Z}}^d \cap \Lambda_N}  \int_{{\mathbb{S}}^{d-1}}dp_i \exp 
\left( \frac{-1}{2} \sum_{1\leq k,j \leq n} (p_k \cdot \partial)(p_j \cdot \partial) C(x_k, x_j)    \right) \\
\end{split}
\end{equation}
All fundamental objects to study, such as the pressure, truncated correlation functions, etc. are derived from $Z_N$ and similar objects.

The model can be equivalently expressed  as a Euclidean field theory (due to Kac \cite{Kac59} and Siegert \cite{Sie60}) and is given by 
 \begin{equation}   \label{SineGordon}
\begin{split}
{_{0}Z}_N \equiv Z_N  &=  \int    \exp  \Big( zW( \Lambda_N, \phi)  \Big)   d\mu_{ C } (\phi) \\
\textrm{where} \hspace{1cm} W( \Lambda_N, \phi)   &=   \sum_{x \in  \Lambda_N \cap  {\mathbb{Z}}^d}       \int_{{\mathbb{S}}^{d-1}}  dp     \cos  (   p  \cdot  \partial \phi(x)  )
\end{split}
\end{equation}
with
\begin{itemize} 
\item  $dp$  : the standard normalized rotation invariant measure on  ${\mathbb{S}}^{d-1}$. 
\item The fields    $\phi(x)$    :  a    family of   Gaussian random  variables (on some abstract measure space)  indexed by  $x \in {\mathbb{Z}}^d$  with mean zero and covariance  $C(x,y)$ which is a  positive  definite function as given above.  
\item The measure   $\mu_C$  : the underlying measure (see section 11.2 \cite{Dim11} and Appendix A \cite{Dim09} for more  detail). We discuss about  the equivalence of (\ref{grand}) and (\ref{SineGordon}) in  appendix \ref{KacSie} of this paper.
\end{itemize}

For investigating the truncated correlation functions, we consider a  more general version of (\ref{SineGordon}): 
 \begin{equation}   \label{partition1}
\begin{split}
{_{f}Z}_N   &=  \int    \exp  \Big(i f (\phi) + zW( \Lambda_N, \phi)  \Big)   d\mu_{ C } (\phi) \\
\end{split}
\end{equation}
where $f(\phi)$ can be:
\begin{enumerate}
\item $f(\phi) = 0$ as in \cite{Dim09}.

\item $f(\phi) = \sum_{k=1}^m t_k \partial_{\mu_k} \phi (x_k)$. We use this $f(\phi)$ to study the truncated correlation functions\\ $\cG^t (x_1, x_2,... x_m)  \equiv  \big<\prod_{k=1}^m \partial_{\mu_k} \phi(x_k)\big>^t =  i^m \frac{\partial^m}{\partial t_1 \dots  \partial t_m} \log {_{f}Z'} \Big|_{t_1=0,\dots t_m=0} $ which is nontrivial and previously investigated by Dimock and Hurd \cite{DimHur92}.

\item $f(\phi) = \sum_{k=1}^m t_k \exp \left( i\partial_{\mu_k} \phi(x_k) \right)$ (for studying  the dipole correlation \\
$ \big<\prod_{k=1}^m \exp \left( i\partial_{\mu_k} \phi(x_k) \right) \big>^t =  i^m \frac{\partial^m}{\partial t_1 \dots  \partial t_m} \log {_{f}Z'} \Big|_{t_1=0,\dots t_m=0} $ ).

\item Other general form which will be discussed at the end of this paper. The general form can be applied for truncated correlations of  density of the dipoles which also has been studied by Brydges and Keller \cite{BryKel93}. We think that this general form has more applications.
\end{enumerate}
Here  $x_k \in {\mathbb{Z}}^d$ are different points; $\mu_k \in \{\pm1,\dots,\pm d \}$ and $t_k$ small and complex; $m \geq 2$. For the set $\{ x_1,\dots,x_m\} \subset {\mathbb{Z}}^d$, let $ \textrm{ diam} (x_1,\dots,x_m) = \max_{1 \leq i,j \leq m} \textrm{dist} (x_i,x_j)$ where
$\textrm{dist} (x_i,x_j)$ is the distance between $x_i$ and $x_j$ on lattice ${\mathbb{Z}}^d$.

To get rid of the boundary and study the long distance properties of the system,  we would like to take the thermodynamic limit for these quantities, i.e. the limit as $N \to \infty$  which is so called \textit{infinite volume limit}. 
Actually  ${Z}_N$   is not expected to have a limit as $N \to \infty$. In \cite{Dim09}, Dimock has  established an infinite volume limit for the pressure defined by    
\begin{equation}
{p}_N   =  | \Lambda_N|  ^{-1}    \log {_{0}{Z}}_N  
\end{equation}
(with $|z|$ sufficiently small). Such infinite volume limits have also been obtained  by   Frohlich and Park   \cite{FroPar78}   and by  Frohlich and Spencer  \cite{FroSpe81}. They used a method of correlation inequalities.

In   this paper,  we continue the study of the long distance properties of the dipole gas model. For long distance (i.e.,  when $|x-y|$ large), the  potential $ \partial_{\mu}  \partial_{\nu}  C(x-y)$ behaves like   $ {{\cal O}}(  |x-y|^{-d})$, that means it is not integrable and we  could  not use the theory of the Mayer expansion to establish   such   results. To overcome this problem, we use the  method of the renormalization group.

 We  follow  particularly  a   Renormalization Group  approach   recently developed   by  Brydges  and  Slade    \cite{Bry07} and  Dimock \cite{Dim09}.  We generalize  Dimock's framework  with an  external field and obtain some estimates on the  correlation functions as in Dimock and Hurd  \cite{DimHur92}, and  Brydges and Keller \cite{BryKel93}. The main result is the existence of the infinite volume limit for correlations functions, which is new.  Earlier work  using RG  approach to the dipole gas can be found in    Gawedski and Kupiainen  \cite{GawKup84},   Brydges and Yau  \cite{BryYau90}.

 Besides the dipole gas papers mentioned  above,  we would like to cite some other  papers on the Coulomb gas in $d=2$ which has a dipole phase.  There are the works of  Dimock and Hurd  \cite{DimHur00}, Falco \cite{Fal12} and Zhao \cite{Zhao}.

\subsection{The main result}  
 For our RG approach  we follow the analysis of    Brydges' lecture   \cite{Bry07}.  Instead of  (\ref{partition1}), we  use   a   different  finite volume  approximation. First, we   add    an extra term   $(1- \varepsilon)V( \Lambda_N,  \phi  )$ where   $0< \varepsilon$ is  closed to 1 and 
\begin{equation}  \label{VLaN}
V(\Lambda_N,  \phi  )  =   \frac{ 1}{4} \sum_{  x  \in  \Lambda_N  \cap  {\mathbb{Z}}^d}  \sum_{\pm \mu =1}^d   
 ( \partial_{\mu} \phi(x) )^2 
 \end{equation}
By replacing the covariance   $C$ by 
 $\varepsilon^{-1} C$, this  extra term will be partially    compensated. Hence  instead of   (\ref{partition1})  we will consider  a new  finite  volume  generating function
\begin{equation}   \label{partition4} 
\begin{split}
{_{f}Z}_N  &=   {_{f}Z}'_N /Z''_N\\
\end{split}
\end{equation}
where
\begin{equation}   \label{partition2} 
\begin{split}
{_{f}Z}'_N &=  \int  e^{i f(\phi)}    \exp  \Big(z W( \Lambda_N,   \phi)-(1-\varepsilon) V(\Lambda_N,  \phi  ) \Big) 
  d\mu_{\varepsilon^{-1} C } (\phi) \\ 
\end{split}
\end{equation}
and
\begin{equation}   \label{partition3} 
Z''_N =  \int      \exp  \Big(-(1- \varepsilon)V( \Lambda_N,  \phi  ) \Big) 
  d\mu_{\varepsilon^{-1} C } (\phi)  
\end{equation}
We have 
\begin{equation}
\begin{split}
{_{f}Z}_N  &=   {_{f}Z}'_N /Z''_N  \\
&= \int  e^{i f(\phi)}    \exp  \left(z W( \Lambda_N,   \phi) \right) \Big[  (Z"_N)^{-1} \exp\left(-(1-\varepsilon) V(\Lambda_N,  \phi  ) \right) 
  d\mu_{\varepsilon^{-1} C } (\phi) \Big]
\end{split}
\end{equation}
When $N \to \infty$,  $ \exp  \left(-(1- \varepsilon)V( \Lambda_N,\phi  ) \right)$  formally becomes $ \exp  \left(  1/2 (1- \varepsilon)(\phi, -\De \phi) \right)$, and $ d\mu_{\varepsilon^{-1} C } (\phi) =  \left(  1/2 (\varepsilon)(\phi, -\De \phi) \right) d \phi$. So the bracketed expression formally converges to $\textrm{(const.)} \left(  1/2 (\phi, -\De \phi) \right) d \phi = \textrm{(const.)}  d \mu_C(\phi)$ when $N \to \infty$. Formally this new ${_{f}Z_N}$ gives   the same limit as   (\ref{partition1}).     This result holds  for any choice of  $\varepsilon$. By definition (\ref{VLaN}), the extra term $(1- \varepsilon)V( \Lambda_N,  \phi  )$    $= (1- \varepsilon) \frac{ 1}{4} \sum_{  x  \in  \Lambda_N  \cap  {\mathbb{Z}}^d}  \sum_{\pm \mu =1}^d    ( \partial_{\mu} \phi(x) )^2 $. Therefore   the choice of $\varepsilon$ is a choice of how  much  $(\partial \phi)^2$  one is putting in the interaction  and how much in the measure .  

\bigskip

Similarly to the Theorem 1 in  \cite{Dim09}, our main theorems are:
\begin{thm} \label{T1}
 For     $|z|$ and $\max_{k} |t_k|$  sufficiently    small  there is an    $\varepsilon  =  \varepsilon (z)$  close  to 1
so  that    \\
 $ {_{f}p}_N  =     | \Lambda_N|  ^{-1}    \log\big(  {_{f}Z}_N   \big) $
has a limit as  $N  \to  \infty$.\footnote{In Theorem \ref{T1}, $f(\phi)$ can be $0$, $\sum_{k=1}^m t_k \partial_{\mu_k} \phi (x_k)$, or  $\sum_{k=1}^m t_k \exp \left( i\partial_{\mu_k} \phi(x_k) \right)$.}   
\end{thm}

Using $f(\phi) = \sum_{k=1}^m t_k \partial_{\mu_k} \phi (x_k)$, we achieve  some  estimate for the correlation functions:
\begin{thm}  \label{corrpa}
 For any small $\epsilon > 0$, with $L, A$ sufficiently large (depending on $\epsilon$), $\eta = \min \{d/2, 2\}$, we have:    
\begin{equation}
\begin{split}
 \Big| \big<\prod_{k=1}^m \partial_{\mu_k} \phi(x_k)\big>^t \Big| \leq \frac{m!}{a^m}  \textrm{ diam}^{-\eta+\epsilon} (x_1,... x_m)  \\ 
\end{split}
\end{equation}
where $a$ depends on $\epsilon, L, A$.
\end{thm}
And we also can obtain the existence of infinite volume limit for correlation functions.
\begin{thm}  \label{inflimcorr} With $L, A$ sufficiently large, the infinite volume limit of truncated correlation function 

$\lim_{N \to \infty}  \big<\prod_{k=1}^m \partial_{\mu_k} \phi(x_k)\big>^t $ exists
\end{thm}
When $d=3$ or $4$, the result in Theorem \ref{corrpa}  looks like the result in \cite{DimHur92}, but here it is obtained with the new method. 

\bigskip

 Using  $f(\phi) = \sum_{k=1}^m t_k \exp \left( i\partial_{\mu_k} \phi(x_k) \right)$, we can obtain Theorems \ref{CorrExp} and \ref{inflimCorrExp} which are similar to Theorems \ref{corrpa} and \ref{inflimcorr}, just with different $f$.

At the end of this paper, we investigate a general form of $f$ and obtain  Theorems \ref{Corrgen} and \ref{inflimCorrgen}. Applying theorem \ref{Corrgen} with a special  $f$ for density of dipoles, we have obtained some estimates for truncated correlation functions of density of dipoles with $(m \geq 2)$ points (Corollary \ref{CorrDens}) instead of only 2 points as Theorem 1.1.2 in \cite{BryKel93}. Then we apply theorem \ref{inflimCorrgen} to  establish the infinite volume limit for  truncated correlation functions of density of dipoles  (Corollary \ref{inflimCorrDens}).

For the  proof of  Theorem \ref{T1}, we will show that,  with a  suitable   choice of  $\varepsilon  =  \varepsilon(z)$,  the density  $ \exp \big(z W-(1-\varepsilon)V\big) $  likely goes  to zero under the renormalization group flow  and leaves  a   measure  like     $\mu_{\varepsilon(z)^{-1} C}$  to describe  the long distance behavior of the system.  Accordingly    $\varepsilon(z)$   can be interpreted  as   a dielectric constant. 

Now we  rewrite the  generating function ${_{f}Z_N}$. First we     scale    $\phi \to   \phi/\sqrt{ \varepsilon}$   and then let      $\sigma   =   \varepsilon^{-1} -1$.    Because     $\varepsilon$ is  closed to 1 , we have $\sigma$ is  near zero. We also have  
\begin{equation}   \label{fmeasure}
\begin{split}
{_{f}Z}'_N(z,\sigma)= & \int  e^{i f(\phi)}    \exp  \Big( zW(\Lambda_N,   \sqrt {1 +\sigma } \phi)-   \sigma V( \Lambda_N, \phi)\Big)   d\mu_{ C } (\phi)  \\
Z''_N(\sigma)= & \int      \exp  \Big(-  \sigma  V( \Lambda_N, \phi))\Big)   d\mu_{ C } (\phi)  \\
{_{f}Z}_N(z,\sigma)=&{_{f}Z}'_N(z,\sigma)/Z''_N(\sigma)  \\
\end{split}
\end{equation}
Then    we need  to   show that  with      $|z|$  sufficiently    small  there is a  (smooth)     $\sigma   =  \sigma (z)$  near zero     
such that,  
\begin{equation}  \label{newsplit}
\begin{split}
 | \Lambda_N|  ^{-1}    \log {_{f}Z}_N(z, \sigma(z)) 
=    | \Lambda_N|  ^{-1}    \log {_{f}Z}'_N(z, \sigma(z))   
- | \Lambda_N|  ^{-1}    \log Z''_N (\sigma(z)) \\
\end{split}
\end{equation}
has   a limit when  $N \to \infty$.  And theorem \ref{T1} is   proved just by putting $\varepsilon(z)  =  (1 +  \sigma(z))^{-1}$ back. Dimock has proved that, for small  real      $\sigma$  with   $|\sigma|<1$, we have  $ | \Lambda_N|  ^{-1}    \log Z''_N(\sigma)$  converges as   $N \to   \infty$ (Theorem 2,  \cite{Dim09}).  Hence we only need to investigate the first term in (\ref{newsplit})

\bigskip

The  paper is    organized as follows:
\begin{itemize} 
\item In  section  \ref{prelimi} we  give  some  general  definitions  on the lattice and its properties. We also give definitions about the norms we use together with their crucial properties and estimates. Then we  define the basic  Renormalization Group transformation as in (\cite{Dim09}).  
\item In   section \ref{ReGr}   we  accomplish  the detailed analysis of the Renormalization Group   transformation to isolate   the leading terms. Then we simplify them for the next scale.
\item  In   section   \ref{stable}    we   study  the RG flow  and   find  the   stable  manifold   $\sigma  =  \sigma(z)$.    
\item In section  \ref{dipole}   we   assemble  the results and prove the infinite volume limit for  $ | \Lambda_N|  ^{-1}    \log {_{f}Z}'_N   $ exists.
\item  Finally in section 6, by  combining all the other estimates, we obtain some estimates for correlation functions and establish the infinite volume limit of  correlation functions.
\end{itemize}

\section{Preliminaries}  \label{prelimi}

In this section, we quote all notations and basic result from Dimock \cite{Dim09}. At the same time, we  introduce some new notations which are useful for this paper.

\subsection{Multiscale decomposition}  \label{MultiDec}
RG  methods  are  based upon a multiscale decomposition of the basic   lattice   covariance $C$  into  a  sequence  of  more controllable  integrals   and  analyze  the effects separately  at each  stage. Especially we  choose  a decomposition into finite range covariances   which is developed by Brydges,  Guadagni, and Mitter  \cite{BrGuMi03}. 
The  decomposition of the lattice covariance $C$ has the form   
\begin{equation}
C (x-y)   =  \sum_{j=1}^{\infty}     \Gamma_j( x-y )
\end  {equation} 
such that
\begin{itemize} 
\item $\Gamma_j(x)  $  is  defined on  $ {\mathbb{Z}}^d$,   is positive semi-definite, and satisfies the finite range property: $\Gamma_j (x)   = 0$   if  $ |x|  \geq  L^j/2$.
\item There is a   constant  $c_0$  independent of  $L$  such that, for all $j,x$, we have
 \begin{equation}
|  \Gamma_j(x)|   \leq   c_0  L^{-(j-1)(d-2)} 
  \end{equation}
 This implies  that the series converges uniformly.
\item There are  constants  $c_{\alpha}$  independent of  $L$  such  that 
\begin{equation} \label{imp}
| \partial^{\alpha}  \Gamma_j(x)|   \leq   
 c_{\alpha}  L^{-(j-1 )(d-2+ |\alpha|) } 
  \end{equation}
 where  $\partial^{\alpha}  =  \prod_{\pm \mu =1}^d  \partial_{\mu}^{\alpha_{\mu}} $  is  a multi-derivative and         $|\alpha|  =  \sum_{\mu}| \alpha_{\mu} |$.  Thus the differentiated series  converges uniformly   to   $ \partial^{\alpha} C$.
\item (Lemma 2, \cite{Dim09})  There are   some constants  $C_{L,\alpha}$  such that 
\begin{equation}     \label{DevCov}
|\partial^{\alpha}C(x)|  \leq  C_{L,\alpha} ( 1 + |x|)^{-d+2- |\alpha|}
\end{equation}
\end{itemize}

For our  RG analysis  we need to break off pieces  of  covariance $C(x-y)$  one at a time.  So we define 
  \begin{equation}
C_k(x-y)   =  \sum_{j=k+1}^{\infty}    \Gamma_j( x-y  )
\end  {equation} 
Hence we have   $ C = C_0$ and
\begin{equation}
\begin{split}
C_k (x-y) &=   C_{k+1}(x-y)  +  \Gamma_{k+1}(x-y)\\
\end{split}
\end{equation}

\subsection{Renormalization Group  Transformation}  \label{RGTrans}

The  generating function  (\ref{fmeasure})  can be rewritten  as
\begin{equation}
\begin{split}
{_{f}Z}'_N(z,\sigma)=&  \int {_{f} \cZ}_0^N  (\phi)      d\mu_{C_0} (\phi) \\
\end{split}
\end{equation}
with  
\begin{equation}  \label{ZIni}
\begin{split}
{_{f}\cZ}_0^N  (\phi)   &= e^{i f(\phi)} \exp  \Big(zW( \Lambda_N,   \sqrt{1+\sigma} \phi)- \sigma  V( \Lambda_N, \phi)\Big)   \\
\end{split}
\end{equation}
We use the left subscript $f$ as an  extra notation for 3 cases at the same time: 
\begin{itemize}
\item $f(\phi)=0$  as in (Dimock, \cite{Dim09}); 
\item $f(\phi) = \sum_{k=1}^m t_k \partial_{\mu_k} \phi (x_k))$; 
\item $f(\phi) = \sum_{k=1}^m t_k \exp \left( i\partial_{\mu_k} \phi(x_k) \right)$.
\end{itemize}
Since   $C_0   = C_1 +  \Gamma_1$  we replace   an  integral over  $\mu_{C_0}$   by  an integral over $\mu_{\Gamma_1}$  and  $\mu_{C_{1}}$. So we have   
\begin{equation}
\begin{split}
{_{f}Z}'_N(z,\sigma)
=&  \int   {_{f}\cZ}^N_{0}(\phi+  \zeta  )  d  \mu_{\Gamma_1} (\zeta)   d \mu_{C_{1}}(\phi)
=  \int    {_{f}\cZ}^N_1(\phi)    d \mu_{C_1}(\phi)\\
\end{split}
\end{equation}
We define  a  new  density   by  the fluctuation integral  
\begin{equation}
 {_{f}\cZ}^N_{1}(\phi)    =   ( \mu_{\Gamma_1 } *  {_{f}\cZ}^N_{0})(\phi)   \equiv   \int     {_{f}\cZ}^N_{0}(\phi  + \zeta  )  d  \mu_{\Gamma_1} (\zeta)
   \end{equation}
Because      $\Gamma_1,  C_1$ are only positive  semi-definite,   these are degenerate Gaussian measures.\footnote{Dimock has  discussed   these  in Appendix A, \cite{Dim09}.}
By continuing this way, we will  have the representation  for $j=0,1,2, \dots  $ 
\begin{equation}  \label{volume}
{_{f}Z}'_N(z,\sigma)=    \int    {_{f}\cZ}^N_{j}(\phi)      d \mu_{C_j}(\phi)
\end{equation}
here  the density   $  {_{f}\cZ}^N_{j}(\phi) $  is  defined   by  
\begin{equation}  
  {_{f}\cZ}^N_{j+1}(\phi)   =   ( \mu_{\Gamma_{j+1}}   * {_{f}\cZ}^N_{j})(\phi)  
   =   \int     {_{f}\cZ}^N_{j}(\phi  + \zeta  )  d  \mu_{\Gamma_{j+1}} (\zeta) 
\end{equation}
 Our job  is to investigate the growth of these densities when  $j$ go to $\infty$.

\subsection{Local expansion}

We will rewrite each density   ${_{f}\cZ}^N_j(\phi)$  in a form  which presents  its locality properties known as
a polymer representation.    The  localization becomes coarser when   $j$  gets bigger. First we will give some basic definitions on the lattice $\mathbb{Z}^d$.

\subsubsection{Basic definitions on  the lattice ${\mathbb{Z}}^d$}

For $j=0,1,2, \dots$   we  partition $ {\mathbb{Z}}^d$  into  \textit{$j$-blocks}   $B$.  These blocks have side  $L^j$  and   are translates   of the center $j$-blocks
\begin{equation}
B_j^0 =  \{ x \in {\mathbb{Z}}^d:   |x|  <  1/2 (L^j-1)\} 
\end{equation}
  by  points in the lattice   $L^j{\mathbb{Z}}^d$.  The set  of all  $j$-blocks in  $\Lambda = \Lambda_N$ is denoted  $\cB_j(\Lambda_N)$, $\cB_j(\Lambda)$ or just  $\cB_j$.  
A union  of  $j$-blocks $X$ is called a \textit{$j$-polymer}. Note that  $\Lambda$  is also  a  {$j$-polymer} for   $0 \leq j  \leq  N$.  The set of all $j$-polymers in $\Lambda = \Lambda_N$  is denoted   $\cP_j(\Lambda)$ or just  $\cP_j$.   The set of all connected $j$-polymers is denoted by $\cP_{j,c}$. Let $X \in \cP_j$, the closure $\bar{X}$ is the smallest $Y \in \cP_{j+1}$ such that $X \subset Y$. 

For a  $j$-polymer $X$, let $|X|_j$ be the number  of  $j$-blocks in $X$.  We call   $j$-polymer $X$  \textit{a small set}     if it is connected  and  contains no more than $2^d$ $j$-blocks.  The  set  of all small set   $j$-polymers in $\Lambda$  is denoted by  $\cS_j(\Lambda)$ or  just  $\cS_j$. 
A  $j$-block  $B$  has a small set neighborhood   $ B^*   =  \cup   \{  Y \in \cS_j:   Y \supset  B \}$.
{\bf Note:} If $B_1, B_2$ are $j$-blocks and $B_2 \in B_1^*$ then, using above definition, we also have that $B_1 \in B_2^*$. Similarly a  $j$-polymer  $X$   has a small set  neighborhood  $X^*$.

For    $l \geq 1$ and integer $d$, we define some constants $n_1(d), n_2(d), n_3(d,l)$ which are bounded  and, for every $j \geq 0$, we have:
\begin{equation} \label{n123}
\begin{split}
n_1 (d) \equiv&  \sum_{X \in \cS_0, X \supset  0} 1/ | X |_0 =  \sum_{X \in \cS_j, X \supset  B_j^0 } 1/ | X |_j \\
n_2 (d) \equiv&  \sum_{X \in \cS_0, X \supset  0} 1  =  \sum_{X \in \cS_j, X \supset  B_j^0 } 1 \\
n_3 (d,l) \equiv&  \sum_{X \in \cS_0, X \supset  0} \frac {l^{- | X |_0}}{| X |_0} =  \sum_{X \in \cS_j, X \supset   B_j^0} \frac {l^{- | X |_j}}{| X |_j} \\
n_3(d,l) \leq& n_3(d,1) = n_1 (d) \leq n_2 (d) \leq (2^d)! (2d)^{2^d}
\end{split}
\end{equation}
Furthermore, with a fixed $d$, we can get 
\begin{equation}\label{n3dl}
0 \leq \lim_{l  \rightarrow  \infty} n_3(d,l) \leq   \lim_{l  \rightarrow  \infty} \frac{n_1(d)}{l} = 0
\end{equation}

\subsubsection{Local expansion}
Using the same approach as in \cite{Dim09}, we rewrite the  density   $   {_{f}\cZ}^N_j(\phi)$   for  $\phi:  {\mathbb{Z}}^d  \to  {\mathbb{R}}$   in the    the general form  
\begin{equation}   
{_{f}\cZ}    =  ({_{f}I}  \circ   {_{f}K} )  ( \Lambda )  \equiv   \sum_{X  \in \cP_j(\Lambda)}    {_{f}I}(\Lambda - X)  {_{f}K}(X)
\end{equation} 
Here   ${_{f}I}(Y)$ is  a background functional  which is   explicitly known  and   carries the main contribution to  the density.    
The  ${_{f}K}(X)$  is  so called  {\em a polymer  activity}. It  represents  small corrections to the  background.

In  section \ref{dipole} we will show that the initial density ${_{f}I_0}$  has the factor property. We want to keep this factor property at  all scales. Then we can use the analysis of Brydges' lecture \cite{Bry07}. Therefore we  assume     ${_{f}I}(Y)$  always is in the  form of
\begin{equation}
{_{f}I}(Y )  =  \prod_{B \in \cB_j:  B \subset Y}   {_{f}I}(B)
\end{equation}
and       ${_{f}I}(B, \phi)$    depends  on  $\phi$   only    $B^*$, the small set neighborhood of $B$.   Moreover we  assume ${_{f}K}(X)$    factors   over  the  connected components $\cC(X)$  of   $X$
\begin{equation}
{_{f}K}(X)   =  \prod_{Y \subset \cC(X)}    {_{f}K}(Y  )
\end{equation}
and   that  ${_{f}K}(X, \phi)$  only depends on $\phi$  in  $X^*$.     

As in \cite{Dim09}, the background  functional ${_{f}I}(B)$    has  a special  form: $ {_{f}I} ({_{f}E}, \sigma ,  B)   =  \exp(-V({_{f}E}, \sigma,  B))$ where 
\footnote{Sums over  $\mu$ are understood to range over  $ \mu  = \pm1, \dots, \pm d$, unless otherwise 
specified.}  
 \begin{equation}   \label{sigma}
  V({_{f}E}, \sigma ,B, \phi )   =      {_{f}E}(B)   
+   \frac {1}{4}  \sum_{x \in B} \sum_{ \mu \nu  }  \sigma_{\mu \nu} (B) \partial_{\mu}  \phi(x)     \partial_{\nu}  \phi(x)      
\end{equation} 
for some functions   ${_{f}E}, \sigma_{\mu \nu}:   \cB_j  \to {\mathbb{R}}$. Indeed we  usually can take   $\sigma_{\mu \nu}(B)  =  \sigma  \de_{\mu \nu}$  for some constant  $\sigma$. Then $ V({_{f}E}, \sigma ,B, \phi )$ becomes
\begin{equation}  \label{41}
 V({_{f}E}, \sigma ,B, \phi )   =     
     {_{f}E}(B)   +   \frac {\sigma}{4}  \sum_{x \in B} \sum_{ \mu}  ( \partial_{\mu}  \phi(x) ) ^2    
     \equiv  {_{f}E}(B)  + \sigma  V(B)
\end{equation}

Also  in our model, when $f=0$,    we  will  have   
\begin{equation}  \label{symmetry}
\begin{split}
{_{0}K}(X, \phi) \hspace{0.1cm}= \hspace{0.1cm} {_{0}K}(X, -  \phi)   \hspace{2cm}
{_{0}K}(X, \phi) \hspace{0.1cm}=  \hspace{0.1cm} {_{0}K}(X, \phi+c)  \\
\end{split}
\end{equation}
The later holds   for any constant  $c$  which means that   ${_{0}K}(X, \phi)$  only depends   on derivatives   $\partial \phi$.

\subsection{About norms and their properties}

In this paper we use exactly the same norms and notations as in Dimock \cite{Dim09}. 
Now we consider potential $ V( s ,B, \phi )$ of the form  
\begin{equation}
 V( s ,B, \phi )   =  \frac 14  \sum_{x \in B} \sum_{\mu \nu} s_{\mu \nu}(x)   \partial_{\mu}  \phi(x)     \partial_{\nu}  \phi(x)       \\
\end{equation}
here  the norms of functions $s_{\mu \nu}(x)$ are defined by  
\begin{equation}     \label{normSmunu}
\|  s \|_j  =      \sup_{B \in \cB_j} |B|^{-1}  \|s\|_{1,B}  =  \sup_{B \in \cB_j} L^{-dj} \sum_{\mu \nu}
\sum_{x \in  B}  |s_{\mu \nu} (x)|
\end{equation}
If    $s_{\mu \nu}(x)  =  \sigma \de_{\mu \nu}$   then    $V(s, B)  = \sigma V(B)$ as defined in  (\ref{41})   and the norm $\|s \|_j =2d\ \sigma$. 

The following lemmas are some results from Section 3 in \cite{Dim09}:
\begin{lem}   { (Lemma 3, \cite{Dim09})  }
\begin{enumerate}
\item   For any  functional $s_{\mu\nu}(x)$, we have
 \begin{equation}    \label{first}
\begin{split}
\|   V(s, B)  \|'_{s,j} &\leq  h^2 \| s \|_j   \\
 \|   V( s, B)  \|_{s,j}      &\leq       h^2 \| s \|_j 
\end{split}
\end{equation}
\item  The function    $\sigma  \to   \exp ( - \sigma V(B))$ is complex
analytic    and    if       $h^2  \sigma   $  is sufficiently small, we have
\begin{equation}  \label{second}
\begin{split}
   \| e^{- \sigma V(  B)} \|'_{s, j}  &\leq  2   \\
     \| e^{- \sigma  V( B)} \|_{s, j}    &\leq   2
\end{split}
\end{equation}
\end{enumerate}
\end{lem}

Let $c$ be a constant such that the function $\sigma \to \exp (- \sigma V(B)  )$ is analytic in $|\sigma| \leq ch^{-2}$ and satisfies 
$\| \exp ( - \sigma V(B)  ) \|_{s,j}\leq 2$  on that domain.  

To start the RG transformation, we  also need some  estimate on the initial interaction.   When $j=0$,  $B \in \cB_0$ is just a single site $x \in {\mathbb{Z}}^d$, so  we consider
\begin{equation}  \label{defWuB}
W(u,B, \phi)   =  %2  
\int_{{\mathbb{S}}^{d-1}}  dp    
  \cos  (   p  \cdot  \partial \phi(x) u )
\end{equation} 
\begin{lem}  \label{WuB} {(Lemma 4, \cite{Dim09})}
  {  \  }
 \begin{enumerate}
\item     $W(u,B)  $  is bounded by
\begin{equation}  
\|W(u, B  )   \|_{s,0}      \leq 2   e^{\sqrt{d}hu}
\end{equation}
We also have that $ W(u,B)$  is strongly  continuously  differentiable in $u$. 
\item    $ e^{zW(u, B)}$   is complex   analytic in $z$  and  satisfies,  for   $|z|$   sufficiently small  (depending on $d,h,u$), we have
\begin{equation} 
\|  e^{zW(u,  B)} \|_{s,0}   \leq     2  
\end{equation}
 And $  e^{zW(u, B)}$  is also  strongly  continuously  differentiable in  $u$.    
\end{enumerate}
\end{lem}

\section{Analysis of the RG  Transformation}  \label{ReGr}

Now we use the Brydges-Slade  RG   analysis and follow the framework of Dimock \cite{Dim09}, but with an external field $f$.

\subsection{Coordinates $({_{f}I}_j, {_{f}K}_j)$}    
Continuing to the subsection 2.3.2 (Local Expansion), we suppose that we    have   $_{f}\cZ( \phi)   =    ({_{f}I}  \circ  {_{f}K})( \Lambda,  \phi)$  with polymers  on scale  $j$. 
We  rewrite it as  
 \begin{equation}  \label{RG}
  _{f}\cZ'(\phi')   =   ( \mu_{\Gamma_{j+1}}   * {_{f}\cZ})(\phi')   \equiv   \int   { _{f}\cZ}(\phi'  + \zeta  )  d  \mu_{\Gamma_{j+1}} (\zeta) 
\end{equation}
here  we try to put it back to the form 
\begin{equation} \label{fZprime}
_{f}\cZ'( \phi')   =    ({_{f}I} ' \circ {_{f} K}')( \Lambda,  \phi') 
\end{equation} 
 where   the polymers are now on scale  $(j+1)$. Furthermore, supposed that  we  have chosen    ${_{f}I}'$,  we will find  ${_{f}K}'$ so the identity holds. As explained before, our choice of    ${_{f}I}'$  is  to have     the   form
\begin{equation}
{_{f}I}'(B' , \phi')   =   \prod_{B \in \cB_j, B \subset  B'}  \tilde {_{f}I}  (B, \phi')          { \hspace{1cm}}     B' \in \cB_{j+1}
\end{equation}
Now we define 
\begin{equation}
\begin{split}
\de {_{f}I} (B, \phi', \zeta)  =&   {_{f}I}(B, \phi' + \zeta)  - \tilde {_{f}I} (B,  \phi')\\
 {_{f}K}  \circ   \de {_{f}I}\equiv \tilde {_{f}K}(X, \phi', \zeta) =&  \sum_{Y \subset X}  {_{f}K}(Y,  \phi' + \zeta)  \de {_{f}I}^{X-Y} ( \phi', \zeta)\\
\end{split}   
  \end{equation}     
 For connected $X$  we write  $\tilde  {_{f}K}(X, \phi', \zeta)$ in the form   \footnote{As in (Dimock, \cite{Dim09}), ${_{f}J}(B,X)$  will   be chosen to depend on  ${_{f}K}$ and required   ${_{f}J}(B,X) =0$ unless  $X  \in \cS_j,  B \subset X$   and that    ${_{f}J}(B,X, \phi')$  depend on  $\phi'$ only in $B^*$.}
\begin{equation}  \label{Kcheck}
\tilde  {_{f}K}(X, \phi', \zeta)  =  \sum_{B \subset  X}   {_{f}J}(B, X, \phi')  +   \check{_{f}K}(X, \phi', \zeta)
\end{equation} 
Given  ${_{f}K}$ and  ${_{f}J}$  the equation (\ref{Kcheck}) would give us a definition of    $\check {_{f}K}(X)$  for $X$ connected.  And for  any   $X \in \cP_j$,   we define
\begin{equation}
\check{_{f}K}(X, \phi', \zeta)  =  \prod_{Y \in \cC(X)}   \check{_{f}K}(Y, \phi',\zeta)
\end{equation}

After using the finite range property and making some rearrangements as Proposition 5.1, Brydges \cite{Bry07}, we have (\ref{fZprime})  holds  with   
\begin{equation}  \label{basic}
{_{f}K}'(U, \phi')  =   \sum_{X,\chi   \to  U}  {_{f}J}^{\chi}(\phi')   \tilde  {_{f}I}^{U -(X_{\chi}  \cup X)}(\phi') \check {_{f}K}^\#(X, \phi') 
{ \hspace{.7cm}} U \in \cP_{j+1}
\end{equation}
where $\chi   =  (B_1, X_1, \dots   B_n, X_n) $  and the  condition  $X,\chi   \to  U$  means  that   $X_1,  \dots  X_n,  X $  be strictly disjoint   and satisfy $\overline{ (B^*_1  \cup  \cdots  \cup B^*_n  \cup  X)}  = U$.   Moreover   
\begin{equation}
\begin{split}
 {_{f}J}^{\chi}(\phi')   = &  \prod_{i=1}^n    {_{f}J}(B_l, X_l, \phi')\\
 \tilde  {_{f}I}^{U -(X_{\chi}  \cup X)} (\phi') =&  \prod_{B \in  U -(X_{\chi}  \cup X)} \tilde   {_{f}I}(B, \phi')\\
\end{split}
\end{equation}
with    $X_{\chi}  =  \cup_i  X_i$.  And   $\check {_{f}K}^\#(X, \phi') $  is  $\check {_{f}K}(X, \phi', \zeta )$
integrated over $\zeta$ as (78) in \cite{Dim09}.

At this point  ${_{f}K}'$  is considered as a function  of  ${_{f}I}, \tilde {_{f}I}, {_{f}J}, {_{f}K}$.  
It  vanishes  at the point $({_{f}I}, \tilde {_{f}I}, {_{f}J}, {_{f}K}) =  (1,1,0,0)$ since   $\chi = \emptyset$  and  $X  = \emptyset$ iff  $U = \emptyset$.
We  study  its behavior in a neighborhood  of this point.  We have  the norm on  ${_{f}K}$  as (75) in \cite{Dim09} and   we  define
\begin{equation}
\begin{split}
 \| {_{f}I}\|_{s,j}   = & \sup_{B  \in \cB_j}\| {_{f}I}(B)\|_{s,j}  \\
 \|\tilde   {_{f}I}  \|'_{s,j}    = & \sup_{B  \in \cB_j} \|\tilde  {_{f}I}(B)\|'_{s,j}  \\
 \| {_{f}J} \|'_j    = & \sup_{X \in \cS_j, B \subset X}\|  {_{f}J}(X,B)\|'_j \\
 \end{split}
\end{equation}
We also set 
\begin{equation} \label{deK}
\de {_{f}K} =  {_{f}K} - {_{0}K}
\end{equation}
Using the same argument as Theorem 3 in \cite{Dim09}, we have  the following result.
 \begin{thm}  Let  $A$ be sufficiently large.  \label{T4}
 \begin{enumerate}
 \item For         $R>0$    there is a $r >0$  such  that the following holds for  all   $j$.    If      $ \|{_{f}I}-1\|_{s,j}< r$,       $\|\tilde {_{f}I}-1\|'_{s,j}< r$,  $\max \{ \| {_{f}J} \|'_j ,  \| {_{0}J} \|'_j \}  < r $  and $ \max \{  \| {_{f}K} \|_j , \| {_{0}K} \|_j \} < r $ 
  then  $\max \{  \| {_{f}K}'\|_{j+1} , \| {_{0}K}'\|_{j+1} \} <   R$.  Furthermore   ${_{f}K}'$  is a smooth function 
 of  ${_{f}I}, \tilde {_{f}I},  {_{f}J}, {_{f}K}$ on this domain with derivatives bounded uniformly in $j$. The analyticity of ${_{f}K}'$ in $t_1,\dots,t_m$ still holds when we go from $j$-scale to $(j+1)$-scale.
 \item    If also  
 \begin{equation}    \label{zero}
\sum_{X \in \cS_j:    X \supset B}  {_{f}J}(B, X)  = 0
\end{equation}
 then the      linearization   of  ${_{f}K}' =  {_{f}K}'({_{f}I}, \tilde {_{f}I}, {_{f}J}, {_{f}K})$   at 
  $({_{f}I}, \tilde {_{f}I}, {_{f}J}, {_{f}K})$  $=(1,1,0,0)$  is  
\begin{equation}
\sum_{\substack{X  \in \cP_{j,c}\\  \overline{X}  = U}}  \left(
{_{f}K}^\#(X)  +  ({_{f}I}^\#(X)-1)1_{X \in \cB_j}  - (\tilde {_{f}I}(X)-1) 1_{X \in \cB_j}  -  \sum_{B \subset X}   {_{f}J}(B,X)\right)
\end{equation} 
where  
\begin{equation}    \label{Ksharp}
{_{f}K}^\#(X, \phi)  = \int  {_{f}K}(X,  \phi+ \zeta) d \mu_{\Gamma_{j+1}} ( \zeta)
\end{equation}
\end{enumerate}
and ${_{0}J}$ actually is ${_{f}J}$ at $f = 0$. 
 \end{thm}

\subsection{Choosing $J$ and Estimating $\cL_1, \cL_2$}
\subsubsection{Choosing $J$}
For a smooth   function $g(\phi)$ on   $\phi  \in  {\mathbb{R}}^{\Lambda}$   let   $T_2g$  denote  a second order
Taylor expansion:
\begin{equation}
\begin{split}
(T_2g)( \phi)   =&  g(0)  + g_1(0; \phi)  + \frac  12  g_2(0; \phi, \phi)    \\
(T_0g)( \phi)   =&  g(0) \\
\end{split}
\end{equation}
With  ${_{f}K}^\#$  defined in  (\ref{Ksharp}), for     $X \in \cS_j$, $ X \supset  B$, $ X \neq B$, we pick:
\begin{equation}  \label{A}
\begin{split}
{_{f}J}(  B,  X) =&    \frac{1}{|X|_j}[ T_2 ( {_{0}K}^\#(X)) + T_0( {_{f}K}^\#(X)) - T_0 ({_{0}K}^\#(X) ) ]   \\
=&    \frac{1}{|X|_j}[ T_2 ( {_{0}K}^\#(X)) + T_0( \de {_{f}K}^\#(X) ) ] \\
\end{split}
\end{equation}
and choose  ${_{f}J}(B, B) $  so that  (\ref{zero}) is satisfied.  Otherwise, we let     ${_{f}J}(B, X)=0$.

As in  (\ref{41})  we have picked   
\begin{equation}   \label{B}
{_{f}I}(B) ={_{f}I}({_{f}E}, \sigma,B) = \exp  (- V({_{f}E}, \sigma, B))  
\end{equation}
  So we require  $\tilde  {_{f}I}$   to  have the  same   form  
 \begin{equation}  \label{C}
 \tilde  {_{f}I}(B) ={_{f}I}(\tilde  {_{f}E}, \tilde  \sigma,B) =  \exp  (- V(\tilde {_{f}E}, \tilde  \sigma, B))    
 \end{equation}
    with     $\tilde {_{f}E},  \tilde \sigma$    which will be defined later. Because  $\sum_{B \subset B'} V(B) = V(B')$, we have 
\begin{equation}
   {_{f}I}'(B') ={_{f}I}( {_{f}E}',   \sigma',B') =  \exp  (- V({_{f}E}', \sigma', B))    
\end{equation}
 with
\begin{equation}
\begin{split}
{_{f}E}'(B') & =  \sum_{B \subset B'} \tilde  {_{f}E}(B) \\
    \sigma' &= \tilde \sigma
\end{split}
\end{equation}

The map      ${_{f}K}'$ becomes  ${_{f}K}' =  {_{f}K}'(\tilde   {_{f}E},\tilde   \sigma, {_{f}E}, \sigma, {_{f}K}, {_{0}K})$. We use the standard norm on the energy
\begin{equation}
\|  {_{f}E}\|_j  =     \sup_{B \in \cB_j} | {_{f}E}(B)| 
\end{equation}
And  the theorem \ref{T4} becomes:
   \begin{thm}  \label{T5} Let  $A$ be sufficiently large. 
 \begin{enumerate}
 \item For    $R>0$    there   is a $r >0$  such that  the following holds for all $j$.  
  If   $\|\tilde  {_{f}E}\|_j$ , $ |\tilde   \sigma|$,  $\|{_{f}E}\|_j  $,  $| \sigma|$, $ \max \{ \| {_{f}K} \|_j , \| {_{0}K} \|_j \} < r$
   then  $\max \{  \| {_{f}K}'\|_{j+1} , \| {_{0}K}'\|_{j+1} \} < R$.  Furthermore   ${_{f}K}'$  is a smooth function 
 of $\tilde  {_{f}E}, \tilde  \sigma, {_{f}E}, \sigma, {_{f}K}, {_{0}K}$    on this domain with derivatives bounded uniformly in $j$. The analyticity of ${_{f}K}'$ in $t_1,\dots,t_m$ still holds when we go from $j$-scale to $(j+1)$-scale.
 \item   
 The      linearization   of  ${_{f}K}'$  at the origin   has the form
 \begin{equation}
 \cL_1( {_{f}K})   +  \cL_2 ({_{f}K})   +  \cL_3({_{f}E}, \sigma, \tilde {_{f}E},  \tilde  \sigma,  {_{f}K}, {_{0}K})
 \end{equation}
 where   
\begin{equation}
\begin{split}
\cL_1 ({_{f}K})(U)  = &    \sum_{\substack{X \in \cP_{j,c}: \\ X \notin \cS_j, \overline X =U}}  {_{f}K}^\#(X)   \\
= &  \sum_{\substack{X \in \cP_{j,c}: \\ X \notin \cS_j, \overline X =U}}  {_{0}K}^\#(X) 
+  \sum_{\substack{X \in \cP_{j,c}: \\ X \notin \cS_j, \overline X =U}} \de {_{f}K}^\#(X) \\
=& \cL_1 ({_{0}K})(U) + \cL_1 (\de{_{f}K})(U) \\
\cL_2 ({_{f}K}) (U)    =&  \sum_{\substack{X \in \cS_j \\ \overline X =U}}  {_{f}K} ^\#(X) - [ T_2 ( {_{0}K}^\#(X)) + T_0( \de {_{f}K} ^\# (X) ) ] \\
=&  \sum_{\substack{X \in \cS_j \\ \overline X =U}}( I - T_2) ( {_{0}K}^\#(X)) 
+ \sum_{\substack{X \in \cS_j \\ \overline X =U}}(I - T_0)( \de {_{f}K} ^\# (X) )  \\
=&   \cL_2 ({_{0}K}) (U)   + \cL_2 (\de{_{f}K}) (U)   \\
%\end{split}
%\end{equation}
%\begin{equation}
%\begin{split}
\cL_3(E, \sigma, \tilde E,  \tilde  \sigma,&  {_{f}K}, {_{0}K})(U)   =   \sum_{\bar B =U}  \left( V (\tilde E, \tilde \sigma, B) - V^{\#} ( E, \sigma, B) \right)\\
&+ \sum_{\bar B =U}  \sum_{\substack{X \in \cS_j \\ X \supset  B}}  \frac{1}{|X|_j} [ T_2 ( {_{0}K}^\#(X)) + T_0( \de {_{f}K} ^\# (X) ) ]\\
\end{split}
\end{equation}
\end{enumerate}
 \end{thm}
  {\noindent{\bf Proof}. }    The new map actually is the composition of the map   $ {_{f}K}' = {_{f}K}'({_{f}I}, \tilde {_{f}I}, {_{f}J}, {_{f}K})$ of theorem \ref{T4}
 with the maps   ${_{f}I} = {_{f}I}( {_{f}E}, \sigma),  \tilde {_{f}I}  =  {_{f}I}(\tilde {_{f}E}, \tilde \sigma),   {_{f}J}  = {_{f}J}( {_{f}K}, {_{0}K})$.  Thus it suffices to establish  uniform bounds and smoothness for the latter.

In the case $f=0$, we've already have the proof in  Theorem 4, Dimock \cite{Dim09}. So we only consider the case $f \neq 0$.
 
For ${_{f}I} = {_{f}I}( {_{f}E}, \sigma),  \tilde {_{f}I}  =  {_{f}I}(\tilde {_{f}E}, \tilde \sigma)$ the proof is the same as the proof for $I, I'$  in (Theorem 4, \cite{Dim09}).  
We have:
\begin{equation}  
\begin{split}
{_{f}J}(  B,  X) =&    \frac{1}{|X|_j}[ T_2 ( {_{0}K}^\#(X)) + T_0( {_{f}K}^\#(X)) - T_0 ({_{0}K}^\#(X) ) ]   \\
  =&    \frac{1}{|X|_j}[( T_2 - T_0) ( {_{0}K}^\#(X)) + T_0( {_{f}K}^\#(X) ) ] \\
\end{split}
\end{equation}
With the same argument as in (Theorem 4, \cite{Dim09}),  %%TuanLe% the technique and argument are the same as in the proof of Lemma 8 , \cite{Dim09} when he's estimated the norms of $H_X$
 we obtain: 
\begin{equation}
\begin{split}
\| ( T_2 - T_0) ( {_{0}K}^\#(X))\|'_j  &\leq  {{\cal O}}(1)  \| {_{0}K}^\#(X)\|'_j \\
\| (  T_0) ( {_{f}K}^\#(X))\|'_j  &\leq  {{\cal O}}(1)  \| {_{f}K}^\#(X)\|'_j \\
\end{split}
\end{equation}
By   (79) in \cite{Dim09}  these are  bounded  by $ {{\cal O}}(1)  [ \| {_{f}K}\|_j  + \| {_{0}K}\|_j  ]$ .  The   same bound holds   for  $\|{_{f}J}(B,B)\|'_j$. Therefore   $\| {_{f}J} \|'_j  \leq    {{\cal O}}(1)  [ \| {_{f}K}\|_j  + \| {_{0}K}\|_j  ]$.

The linearization   is just a computation.  Indeed   ${_{f}J}(B,X)$ is designed so that  
\begin{equation}
\begin{split}
\sum_{X \in \cS_j, \overline{X}  = U}  &\left(
{_{f}K}^\#(X)  -  \sum_{ B \subset X}   {_{f}J}(B,X)\right)\\
 =&  \sum_{ \overline{B}  = U}  \sum_{X \in \cS_j, X \supset  B } \frac{1}{|X|_j} [ T_2 ( {_{0}K}^\#(X)) + T_0( \de {_{f}K} ^\# (X) ) ]   \\
& + \sum_{X \in \cS_j, \overline{X}  = U}    {_{f}K} ^\#(X) - [ T_2 ( {_{0}K}^\#(X)) + T_0( \de {_{f}K} ^\# (X) ) ] \\
 \end{split}  
\end{equation}    
which accounts for the presence of these terms.   Also   the linearization of  $ ({_{f}I}^\#(B)-1)$  is 
$-  V^\#({_{f}E}, \sigma, B)$,  and so forth.
This completes the proof.

\subsubsection{ Estimating $\cL_1, \cL_2$ - the first two linearization parts}
Next  we   make some estimates on the linearization's parts.  First we estimate $\cL_1$ which is the linearization on the large $j$-polymers

\begin{lem}   \label{L1}
Let  $A$ be sufficiently large depending on  $L$.  Then the  operator  $\cL_1$  is a  contraction with a norm which goes to zero as  $A \to \infty$.
 \end{lem}

{\noindent{\bf Proof}. }We use the same proof as in (Dimock, \cite{Dim09}, Lemma 5), but with updated notations. We   estimate by (79) and (80) in \cite{Dim09}
\begin{equation}
\begin{split}
\| \cL_1 ( {_{0}K}) (U)  \|_{j+1}    \leq   \| \cL_1 ({_{0}K}) (U)  \|'_j  
&\leq    \sum_{X \notin \cS_j, \overline X =U} \|   {_{0}K}^\#(X) \|_j' \\
&\leq    \sum_{X \notin \cS_j, \overline X =U}\left( A/2\right) ^{-|X|_j}  \|  {_{0}K} \|_j \\
\end{split}
\end{equation}
and
\begin{equation}
\begin{split}
\| \cL_1 (\de {_{f}K}) (U)  \|_{j+1}    \leq   \| \cL_1 (\de{_{f}K}) (U)  \|'_j  
&\leq    \sum_{X \notin \cS_j, \overline X =U} \|  \de {_{f}K}^\#(X) \|_j' \\
&\leq    \sum_{X \notin \cS_j, \overline X =U}\left( A/2\right) ^{-|X|_j}  \| \de {_{f}K} \|_j \\
\end{split}
\end{equation}
Multiplying    by   $A^{|U|_{j+1}}$  then taking the supremum over   $U$, these  yield
\begin{equation}
\begin{split}
\| \cL_1 ( {_{0}K})  \|_{j+1}   &\leq \left[ 
 \sup_{U}    A^{|U|_{j+1}}   \sum_{X \notin \cS, \overline X =U}\left(  A/2 \right) ^{-|X|_j}  \right] \| {_{0}K}\|_j\\
\| \cL_1 (\de {_{f}K})  \|_{j+1}   &\leq \left[ 
 \sup_{U}    A^{|U|_{j+1}}   \sum_{X \notin \cS, \overline X =U}\left(  A/2 \right) ^{-|X|_j}  \right] \| \de {_{f}K}\|_j\\
\end{split}
\end{equation}
Using lemma 6.18 \cite{Bry07},  the bracketed expression goes to zero as $A \to \infty$.   And we also have
\begin{equation}
\begin{split}
\| \cL_1 ( {_{f}K})  \|_{j+1}  &\leq \| \cL_1 ( {_{0}K})  \|_{j+1}  + \| \cL_1 ( \de{_{f}K})  \|_{j+1}\\
\| \de {_{f}K}\|_j &\leq \| {_{0}K}\|_j + \|  {_{f}K}\|_j\\
\end{split} 
\end{equation}   
Hence for $A$ sufficiently large $\| \cL_1 ( {_{f}K})  \|_{j+1}$ is arbitrarily small.  The idea of lemma 6.18 \cite{Bry07} is that, for  large  polymers  $X$ such that  $\bar X = U$,  the quantity  $|X|_j$ must be much larger than  $|U|_{j+1}$.   

(Q.E.D)

\bigskip

Now we estimate and find an explicit upper bound for $\cL_2$
\begin{lem}   \label{L2}
Let  $L$ be sufficiently large .  Then 
the  operator  $\cL_2$  is a  contraction with a norm which goes to zero as  $L \to \infty$.
 \end{lem}

{\noindent{\bf Proof}. } For $f=0$, we have the Lemma 6, in \cite{Dim09}.

For $f \neq 0$, we can write  
\begin{equation}
\begin{split}
\cL_2 ({_{f}K}) (U) &= \sum_{X \in \cS_j, \overline X =U} \{ (I - T_2) {_{0}K}^\#(X) +  (I - T_0)\de {_{f}K}^\#(X)  \}  \\
&= \cL_2 ({_{0}K}) (U) + \cL_2 (\de {_{f}K}) (U)\\
\end{split}
\end{equation}
where 
\begin{equation}
 \cL_2 (\de {_{f}K}) (U) =   \sum_{X \in \cS_j, \overline X =U}   (I - T_0)\de {_{f}K}^\#(X)  
\end{equation}
Using  (\cite{Bry07}, (6.40)) as well as (\cite{Dim09}, Lemma 6) we get:
\begin{equation}
\begin{split}
 \| (I - T_2) {_{0}K}^\#(X,\phi) \|_{j+1} & \leq   \left(  1 +  \| \phi \|_{\Phi_{j+1}(X^*)} \right)^3  \|  {_{0}K}_3^\#(X, \phi)\|_{j+1} \\
& \leq 4  \left(  1 +  \| \phi \|^3_{\Phi_{j+1}(X^*)} \right)  \|  {_{0}K}_3^\#(X, \phi)\|_{j+1} \\
\| (I - T_0)\de {_{f}K}^\#(X,\phi) \|_{j+1} &  \leq   \left(  1 +  \| \phi \|_{\Phi_{j+1}(X^*)} \right)   \| \de {_{f}K}^\#_1(X,\phi) \|_{j+1} \\
\end{split}
 \end{equation}
Notice that $\de {_{f}K}^\#(X,0) = {_{f}K}^\#(X,0) - {_{0}K}^\#(X,0)$ only depend on $\phi$  in $X^*$. Moreover, ${_{0}K}$ and ${_{f}K}$ are different only on ${ \mathrm{ supp  }} (f) = \{ x_1,..., x_m \}$.\ So, if $X^* \cap \{x_1, x_2,... x_m\} = \emptyset$ then ${_{f}K}^\#(X,0) = {_{0}K}^\#(X,0)$ which means $\de {_{f}K}^\#(X,0) = 0$. Therefore  $\de {_{f}K}^\#(X,0) = 0$ unless  $X^* \cap \{x_1, x_2,... x_m\} \neq \emptyset$

Using property   (64) in \cite{Dim09},  we  have
 \begin{equation}
 \begin{split}
\| {_{0}K}_3^\#(X, \phi)\|_{j+1} &\leq   L^{-3d/2}   \|  {_{0}K}_3^\#(X, \phi)\|_{j} \\ 
& \leq 6 (L^{-3d/2})      \|  {_{0}K}^\#(X, \phi)\|_{j}\\  
&\leq 6 ( L^{-3d/2} )     \|  {_{0}K}^\#(X)\|'_{j}G_j(X,\phi,0)\\
%\end{split}
%\end{equation}
% \begin{equation}
% \begin{split}
 \| \de {_{f}K}^\#_1(X,\phi) \|_{j+1} &\leq  L^{-d/2}   \| \de {_{f}K}^\#_1(X,\phi) \|_{j} \\  
&\leq  L^{-d/2}   \| \de {_{f}K}^\#(X,\phi) \|_{j}\\ 
&\leq  L^{-d/2}   \left(   \| \de {_{f}K}^\#(X)\|_{j}' \right)G_j(X,\phi,0)\\
\end{split}
\end{equation}
and   for  $\phi = \phi'+ \zeta$ , using   (\cite{Bry07}, (6.58)) we get:
\begin{equation}
\begin{split}
 \left(  1 +  \| \phi \|_{\Phi_{j+1}(X^*)} \right) G_j(X, \phi,0)  &\leq  \left(  1 +  \| \phi \|_{\Phi_{j+1}(X^*)} \right)^3 G_j(X, \phi,0) \\ 
&\leq 4 \left(  1 +  \| \phi\|^3_{\Phi_{j+1}(X^*)} \right)
G_j(X, \phi,0)\\
& \leq  4q   G_{j+1}(\bar X,  \phi',\zeta)
\end{split}
\end{equation}
with $q$ as in (\cite{Bry07}, (6.127)). Combining all of them   yields 
\begin{equation}
\begin{split}
  \|(I - T_2) {_{0}K}^\#(X,\phi) \|_{j+1}  &\leq 24q( L^{-3d/2}   )  \| {_{0}K}^\#(X)\|'_{j}  G_{j+1}(\bar X,  \phi',\zeta)\\
\| (I - T_0)\de {_{f}K}^\#(X,\phi) \|_{j+1} &  \leq 4q L^{-d/2} \left(   \| \de {_{f}K}^\#(X)\|_{j}' \right) G_{j+1}(\bar X,  \phi',\zeta)\\
\end{split}
\end{equation}
and    also  using (79) in \cite{Dim09}, we obtain:
\begin{equation}
\begin{split}
\|(I - T_2) {_{0}K}^\#(X)\|_{j+1}   \leq&  24q( L^{-3d/2}   )  \| {_{0}K}^\#(X)\|'_{j} \\
\leq&  24q( L^{-3d/2}   )   ( A/2 )  ^{-|X|_j} \| {_{0}K}\|_{j}\\
\|(I - T_0)\de {_{f}K}^\#(X)\|_{j+1}   \leq& 4q  L^{-d/2}   \| \de {_{f}K}^\#(X)\|_{j}'\\
 \leq& 4q  L^{-d/2}   \left(  \| \de {_{f}K}\|_{j} \right)  A^{-|X|_j} 2^{|X|_j}\\
\end{split}
\end{equation}
Therefore
\begin{equation}
\begin{split}
\|\cL_2 ({_{f}K}) \|_{j+1}   \leq & \|\cL_2 ({_{0}K}) \|_{j+1} + \|\cL_2 (\de{_{f}K}) \|_{j+1}   \\\
 \leq &  24q( L^{-3d/2}   ) \left[ \sup_U A^{|U|_{j+1}}   \sum_{X \in \cS_j, \bar X  = U} ( A/2 )  ^{-|X|_j} \right] \| {_{0}K}\|_{j}\\
&+4q L^{-d/2} \left( \| \de {_{f}K}\|_{j} \right)  \sup_U   \sum_{\substack{X \in \cS_j, \bar X  = U \\ X^* \cap \{x_1, x_2,... x_m\} \neq \emptyset}}                                    A^{|U|_{j+1}}  A^{-|X|_j} 2^{|X|_j}\\
\end{split}
\end{equation}
The bracketed expression  is less than  $2^d 2^{2^d}n_2(d) L^d$  ( using \cite{Bry07}, (6.90))
  so  we have
\begin{equation}
\|\cL_2 ( {_{0}K}) \|_{j+1}   \leq  24q2^2 2^{2^d}n_2(d)( L^{-d/2}   )  \| {_{0}K}\|_{j}   
\end{equation}
Because $|U|_{j+1} \leq |X|_j \leq 2^d$, we get:

\begin{equation} 
\begin{split}
4qL^{-d/2} &\left( \| \de {_{f}K}\|_{j} \right)  \sup_U   \sum_{\substack{X \in \cS_j, \bar X  = U \\ X^* \cap \{x_1, x_2,... x_m\} \neq \emptyset}}                                    A^{|U|_{j+1}}  A^{-|X|_j} 2^{|X|_j}\\
& \leq  4q L^{-d/2} \left( \| \de {_{f}K}\|_{j} \right)  \sum_{\substack{X \in \cS_j, \\ X^* \cap \{x_1, x_2,... x_m\} \neq \emptyset}}                                    2^{2^d}\\
& \leq   4q L^{-d/2} 2^{2^d} \left( \| \de {_{f}K}\|_{j} \right) \sum_{i=1}^m  \sum_{\substack{X \in \cS_j, \\ X^* \cap \{x_i\} \neq \emptyset}}      1                              \\
& \leq 4q m  L^{-d/2} 2^{2^d}  \left( \| \de {_{f}K}\|_{j} \right)  n_2(d)
\end{split}
\end{equation}
Thus
 \begin{equation}
\|\cL_2 ( {_{f}K}) \|_{j+1}   \leq  24q2^2 2^{2^d}n_2(d)( L^{-d/2}   )  \| {_{0}K}\|_{j}   +  L^{-d/2} \big( \| \de {_{f}K}\|_{j}\big)  n_2(d) 2^{2^d}4qm
\end{equation}
Moreover $ \| \de {_{f}K}\|_{j}  \leq \|  {_{0}K}\|_{j}   + \|  {_{f}K}\|_{j}$. So we have the Lemma \ref{L2}. 

(Q.E.D)

\subsection{Splitting $\cL_3$}
\subsubsection{Splitting $\cL_3$}
Similarly in (\cite{Dim09}), we have a special treatment for the term  $\cL_3$.   First  we  rewrite  the final term in  $\cL_3$  which is      
\begin{equation}
\begin{split}
&\sum_{\bar B =U}  \sum_{\substack{X \in \cS_j \\ X \supset  B}}  \frac{1}{|X|_j} [ T_2 ( {_{0}K}^\#(X)) + T_0( \de {_{f}K} ^\# (X) ) ]\\
=&\sum_{\overline{ B} =U}     \sum_{\substack{X \in \cS_j  \\ X \supset   B }} \frac{1}{|X|_j} \left( {_{0}K}^\#(X,0) 
  +  \frac12  {_{0}K}^\#_2(X,0;  \phi,  \phi)  + {_{f}K}^\#(X,0) - {_{0}K}^\#(X,0)\right)\\
=& \sum_{\overline{ B} =U}      \sum_{\substack{X \in \cS_j  \\ X \supset   B }}  \frac{1}{|X|_j} \left( {_{f}K}^\#(X,0) 
  +  \frac12  {_{0}K}^\#_2(X,0;  \phi,  \phi)  \right)\\
\end{split}
\end{equation}
In    $ {_{0}K}^\#_2(X,0;  \phi,  \phi)$  we pick a point  $z  \in   B$, then use the same argument as section 4.3 \cite{Dim09} by replacing	  $\phi(x)$  with  \footnote{We need  the factor $1/2$ since the  sum is over  $\pm \mu = 1, \dots, d$ and    $x_{-\mu} = - x_{\mu} $ }
\begin{equation} \label{86}
    \phi(z) + \frac12  ( x - z)  \cdot  \partial    \phi(z)\
 \equiv  \  \phi(z) + \frac12  \sum_{\mu} ( x_{\mu} - z_{\mu})  \partial_{\mu} \phi(z)
   \end{equation}
 If we also average over     $z \in B$ (\ref{86}) becomes
\begin{equation}  \label{SplitL3}
\begin{split}
&\sum_{\overline{ B} =U}     \sum_{\substack{X \in \cS_j  \\ X \supset   B }}  \frac{1}{|X|_j} \left( {_{f}K}^\#(X,0) 
  +  \frac12  \frac {1}{|B|  }  \sum_{z  \in B} {_{0}K}^\#_2(X,0;  \phi,  \phi)  \right)\\
=& \sum_{\overline{ B} =U}     \sum_{\substack{X \in \cS_j  \\ X \supset   B }} \frac{1}{|X|_j} \left( {_{f}K}^\#(X,0)  \right)\\
&+\sum_{\overline{ B} =U}     \sum_{\substack{X \in \cS_j  \\ X \supset   B }} \frac{1}{|X|_j} \left(  \frac {1}{8 |B|  }  \sum_{z  \in B}\sum_{\mu \nu} {_{0}K}^\#_2(X,0;  x_{\mu}, x_{\nu} )  \partial_{\mu}\phi(z) \partial_{\nu}\phi(z)\right)\\
&+ \sum_{\overline{ B} =U}    \sum_{\substack{X  \in \cS_j:\\ X  \supset  B}}
\frac{1}{|X|_j}   \sum_{z \in B}  \frac{1}{|B|}     \left( \frac12   {_{0}K}^\#_2(X, 0 ;  \phi, \phi) \right)\\
&- \sum_{\overline{ B} =U}     \sum_{\substack{X \in \cS_j  \\ X \supset   B }} \frac{1}{|X|_j} \left(  \frac {1}{8 |B|  }  \sum_{z  \in B}\sum_{\mu \nu} {_{0}K}^\#_2(X,0;  x_{\mu}, x_{\nu} )  \partial_{\mu}\phi(z) \partial_{\nu}\phi(z)\right)\\
%\end{split}
%\end{equation}
%\begin{equation}  \label{SplitL3}
%\begin{split}
=& \sum_{\overline{ B} =U}      \sum_{\substack{X \in \cS_j  \\ X \supset   B }}\frac{1}{|X|_j} \left({_{f}K}^\#(X,0) 
  +   \frac {1}{8 |B|  }  \sum_{z  \in B}\sum_{\mu \nu} {_{0}K}^\#_2(X,0;  x_{\mu}, x_{\nu} )  \partial_{\mu}\phi(z) \partial_{\nu}\phi(z)\right)\\
&  +\cL_3'( {_{f}K}) (U)\\
\end{split}
\end{equation}
here    $\cL_3' ({_{f}K}) (U) = \cL_3' ({_{0}K}) (U) $  is so called the error, namely
 \begin{equation}
\begin{split}
\cL_3' ({_{0}K}) (U)   =  \sum_{\bar B  = U}   \sum_{\substack{X  \in \cS_j:\\ X  \supset  B}}
\frac{1}{|X|_j}   \sum_{z \in B}  \frac{1}{|B|}     \Big( \frac12 &  {_{0}K}^\#_2(X, 0 ;  \phi, \phi)
-  \frac18  {_{0}K}^\#_2(X, 0 ; x \cdot \partial   \phi(z), x \cdot \partial \phi(z))  \Big)\\
 \end{split}
\end{equation}
and we can say $\cL_3' (\de{_{f}K}) (U) = 0$ .  Next    we  define 
\begin{equation}  \label{DefAlBe}
\begin{split}
{_{f}\beta}(B)  =  {_{f}\beta}({_{f}K},B) = & -   \sum_{\substack{X \in \cS_j  \\ X \supset   B }}  \frac{1}{|X|_j}  {_{f}K}^\#(X,0)  \\
\alpha_{\mu \nu}(B) =  \alpha_{\mu \nu}({_{f}K},B)=  \alpha_{\mu \nu}({_{0}K},B)  =&     
-  \frac12  \frac{1}{|B|}  \sum_{\substack{X \in \cS_j  \\ X \supset   B }}  \frac{1}{|X|_j} {_{0}K}^\#_2(X,0; x_{\mu},  x_{\nu})\\
\end{split}
\end{equation}
Note that   $\alpha_{\mu \nu}$ is symmetric and satisfies  $\alpha_{-\mu \nu} =  - \alpha_{\mu \nu}$. We  also let   $\alpha_{\mu\nu}$ stand for  the function $\alpha_{\mu\nu}(x)$  which  takes the constant  value  $\alpha_{\mu \nu}(B)$ for   $x \in B$.

Now we  write   (\ref{SplitL3}) as  
\begin{equation}
\begin{split}
&\sum_{\overline{ B} =U}      \sum_{\substack{X  \in \cS_j\\ X  \supset  B}} \frac{1}{|X|_j} \left({_{f}K}^\#(X,0) 
  +   \frac {1}{8 |B|  }  \sum_{z  \in B}\sum_{\mu \nu} {_{0}K}^\#_2(X,0;  x_{\mu}, x_{\nu} )  \partial_{\mu}\phi(z) \partial_{\nu}\phi(z)\right)
  +\cL_3'( {_{0}K})(U)\\
=&\sum_{\overline{B}  = U} \left(    \frac  14     \sum_{z \in B}  \sum_{\mu \nu} \frac12  \frac{1}{|B|}  \sum_{\substack{X  \in \cS_j\\ X  \supset  B}} \frac{1}{|X|_j} {_{0}K}^\#_2(X,0; x_{\mu},  x_{\nu}) (B) \partial_{\mu} \phi(z) \partial_{\nu}  \phi(z) \right)  \\
&  +      \sum_{\overline{B}  = U}  \sum_{\substack{X  \in \cS_j\\ X  \supset  B}} \frac{1}{|X|_j}  {_{f}K}^\#(X,0)                                                                         + \cL_3' ({_{0}K})(U) \\
\end{split}
\end{equation}
\begin{equation} 
\begin{split}
=& -\sum_{\overline{B}  = U} \left( {_{f}\beta}(B) +   \frac  14     \sum_{z \in B}  \sum_{\mu \nu} \alpha_{\mu \nu} (B) \partial_{\mu} \phi(z) \partial_{\nu}  \phi(z) \right)  +\cL_3' ({_{0}K})(U) \\
=&  - \left( \sum_{\overline{B}  = U}   V({_{f}\beta}, \alpha, B, \phi) \right) + \cL_3' ({_{0}K})(U) \\
\end{split}
\end{equation}
where   $  V({_{f}\beta}, \alpha, B, \phi) $   defined   in  (\ref{sigma}).  Combining all of the above, we get: 
\begin{equation}  \label{SplitL3prime}
\begin{split}
 \cL_3({_{f}E}, \sigma, \tilde  {_{f}E}, \tilde  \sigma,  {_{f}K}, {_{0}K})(U) =  & \sum_{ \overline{B}  = U} 
  \left( V(\tilde {_{f}E}, \tilde  \sigma, B)-V^\#({_{f}E}, \sigma, B)  -    V( {_{f}\beta},  \alpha,  B)  \right)   
 +  \cL'_3 ( {_{0}K}) (U)\\
\end{split}
  \end{equation}

 \subsubsection{Estimating $\alpha, {_{f}\beta}$ and $\cL_3'$}
  First we find some explicit upper bounds for  $\alpha$ and ${_{f}\beta}$
\begin{lem}   \label{albeta}
(Estimates $_{f}\beta$ and $\alpha$)
\begin{equation}
\begin{split}
\|{_{f}\beta} \|_j \equiv  \sup_{B \in \cB_j}  | {_{f}\beta}(B)| \leq & 2 n_2(d) A^{-1} \|{_{f}K}\|_j\\
\|   \alpha  \|_j \equiv    \sup_{B \in \cB_j}  \sum_{\mu \nu}  | \alpha_{\mu \nu}(B) |   \leq & 4 (2d)^2 n_2(d) h^{-2} A^{-1} \| {_{0}K}\|_j\\
\end{split}
\end{equation}
\end{lem}
%\bigskip

{\noindent{\bf Remark}. } The norm  $ \| \alpha \|_j$  agrees with  the norm  $\| s \|_j$ in  (\ref{normSmunu}) if   $s_{\mu \nu}(x)
=  \alpha_{\mu \nu} (B)$ for   $x \in B$.
\bigskip

{\noindent{\bf Proof}. }  By (70) and (79) in \cite{Dim09}, with $A$ very large, we have:  
\begin{equation}  \label{93}
\begin{split}
|  {_{f}K}^\#(X,0)| 
 \leq  &  \| {_{f}K}^\#(X) \|'_j    \leq   (A/2)^{-1}  \| {_{f}K}\|_j \\
\| {_{0}K}_2^\#(X,0)\|_j
\leq &  2 \|   {_{0}K}^\#(X) \|'_j  \leq 4   A^{-1}  \|{_{0}K}\|_j\\
\end{split}
\end{equation}  
From (\ref{n123}), the number of  small sets containing a block $B$  is $n_2(d)$ which is bounded and depends  only 
on $d$, we have:    
\begin{equation}
\begin{split}
|{_{f}\beta}(B)|  &\leq     \sum_{X \in \cS_j, X  \supset B}  | {_{f}K}^\#(X, 0) |  \\
 &\leq     \sum_{X \in \cS_j, X  \supset B}  2 A^{-1} \| {_{f}K}\|_j  \\
&\leq 2 n_2(d)  A^{-1}  \| {_{f}K}\|_j
\end{split}
\end{equation}
We also have   $\| x_{\mu}\|_{\Phi_j(X^*)} =  h^{-1}L^{dj/2}$ and $|B| = L^{dj}$.  By using (66) in \cite{Dim09} we get
\begin{equation}
\begin{split}
|B|^{-1} |  {_{0}K}_2^\#(X,0;  x_{\mu} , x_{\nu})|  & \leq  ( h^{-1}L^{dj/2})^2 L^{-dj}   \|   {_{0}K}_2^\#(X,0) \|_j \\
& =  h^{-2}\|   {_{0}K}_2^\#(X,0) \|_j \\
&\leq   4  h^{-2}  A^{-1}  \| {_{0}K}\|_j \\
\end{split}
\end{equation}
then 
\begin{equation}
\begin{split}
  \sum_{\mu \nu}     |\alpha_{\mu \nu}(B) |
&\leq    \sum_{\mu \nu}   \sum_{X \in \cS_j,X  \supset B}  |B|^{-1}|  {_{0}K}_2^\#(X,0;   x_{\mu},  x_{\nu})|  \\
&\leq    \sum_{\mu \nu}   \sum_{X \in \cS_j,X  \supset B}   4  h^{-2}  A^{-1}  \| {_{0}K}\|_j\\
&\leq    \sum_{\mu \nu}   n_2(d)   4  h^{-2}  A^{-1}  \| {_{0}K}\|_j \\
&\leq    (2d)^2 n_2(d) 4h^{-2} A^{-1}  \| {_{0}K} \|_j
\end{split}
\end{equation}

Now we give some estimate for $\cL_3'$
\begin{lem}    \label{L3prime} 
Let  $L$ be sufficiently large.   Then 
the  operator  $\cL'_3$  is a  contraction with arbitrarily small norm. 
\begin{equation} \label{8new}
\| \cL_3' ({_{0}K})\|_{j+1} \leq 72 d^2 2^{2d}n_1(d)( L^{-2})  \| {_{0}K} \|_j
\end{equation}
 \end{lem}
\bigskip

{\noindent{\bf Proof}. } Based on the proof of  Lemma 8 in \cite{Dim09}, we make some modifications and obtain a better upper bound with some explicit coefficient.

We  have  
 \begin{equation}
\begin{split}
\cL_3'({_{0}K})(U)   =  & \sum_{\bar B  = U}    \sum_{\substack{X  \in \cS_j\\ X  \supset  B}}
\frac{1}{|X|_j}   \sum_{z \in B}  \frac{1}{|B|}      \frac12   {_{0}K}^\#_2(X, 0 ;  \phi- \frac12  x \cdot   \partial \phi(z),  \phi +  \frac12  x \cdot   \partial \phi(z))  \\
\end{split}
\end{equation}
 Using (\cite{Dim09}, (152)- (154)) we get:
\begin{equation} 
\begin{split}
\|   \phi-\frac12   x \cdot   \partial \phi(z)  \|_{\Phi_j(X^*)}  
   \leq &  3d 2^d(   L^{-d/2-1} ) \|  \phi \|_{\Phi_{j+1} (X^*) } \\
\|   \phi+\frac12   x \cdot   \partial \phi(z)  \|_{\Phi_j(X^*)}  
   \leq &  3d 2^d(   L^{-d/2-1} ) \|  \phi \|_{\Phi_{j+1} (X^*) } \\
\end{split}
\end{equation}
Now we  estimate 
\begin{equation}
{_{0}H}_X(U,  \phi)  =    {_{0}K}^\#_2(X, 0 ;  \phi- \frac12  x \cdot   \partial \phi(z),  \phi + \frac12  x \cdot   \partial \phi(z)) 
\end{equation}
Using the same argument as (156)-(157) in \cite{Dim09}, we obtain:
\begin{equation}
\|{_{0}H}_X(U, \phi)\|_{j+1}    \leq     18d^2 2^{2d}( L^{-d-2}) \| K_2^\#(X,0)\|_j (1 + \| \phi\|_{\Phi_{j+1}(U^*)}^2)  \\
 \end{equation}
 But  for  $\phi  = \phi' + \zeta$
 \begin{equation}
 (1 + \| \phi\|_{\Phi_{j+1}(U^*)}^2)   \leq  G_{s,j+1}(U, \phi,0)   \leq  G_{s,j+1}^2(U, \phi',\zeta)  
 \leq    G_{j+1}(U, \phi',\zeta)  
\end{equation}
 Also using     (\ref{93})   we can get:
 \begin{equation}
\| H_X(U)  \|_{j+1}    \leq    72 d^2 2^{2d}( L^{-d-2})  A^{-1} \|K \|_j
 \end{equation}
which yields to
\begin{equation}
\begin{split}
\| \cL'_3K(U)  \|_{j+1}  &\leq    n_1(d)   \sum_{\bar B  = U}  \| H_X(U)  \|_{j+1}  \\
&\leq    n_1(d) L^d   72 d^2 2^{2d}( L^{-d-2})  A^{-1} \|K \|_j \\
&\leq     72 d^2 2^{2d}n_1(d)( L^{-2})  A^{-1} \|K \|_j
\end{split}
\end{equation}
Since   $ \cL'_3K(U)$ is zero  unless   $|U|_{j+1}  =1$  this gives  
\begin{equation}
\| \cL'_3K  \|_{j+1}  \leq       72 d^2 2^{2d}n_1(d)( L^{-2})   \|K \|_j
\end{equation}
(Q.E.D)

\subsection{Identifying  invariant parts and estimating the others }

N0w we investigate the 1st term of  (\ref{SplitL3prime}). We notice that $\alpha_{\mu \nu}(B) =  \alpha_{\mu \nu}({_{f}K},B)=  \alpha_{\mu \nu}({_{0}K},B) $ is independent from $f(\phi)$ and $ {_{0}E}(B),  {_{0}K}(X, \phi)$ actually is the same as $ {E}(B),  {K}(X, \phi)$ in lemma 9 \cite{Dim09}. Therefore we have the same result as lemma 9 \cite{Dim09}

\begin{lem} {(Lemma 9, Dimock \cite{Dim09})}

Suppose   $ {_{0}E}(B),  {_{0}K}(X, \phi)$   are invariant under lattice symmetries   away from the boundary of  $\Lambda_N$
and   $\tilde {_{0}E}(B)$  is  invariant for  $B^*$  away from the boundary. Then 
 \begin{enumerate}
\item   
$ {_{0}E}'(B'),  {_{0}K}'(U, \phi)$  are invariant for  $B', U$   away  from the boundary
\item     If   $B^*$  is away from the boundary then   
${_{0}\beta}(B),\alpha_{\mu \nu}(B) $   are      independent of  $B$   and   $\alpha_{\mu \nu}(B)  =
\hat   \alpha_{\mu \nu}(B)$
defined for all  $B$ by 
 \begin{equation}     \label{defsymalpha}
\hat  \alpha_{\mu \nu}(B)   = \frac{\alpha}{2}  \   (  \de_{\mu \nu}  -  \de_{\mu, -\nu} )
\end{equation}
where   $\alpha$ is a constant.
\end{enumerate}
\end{lem}

\bigskip

  For all  $B \in \cB_j$ we  define
\begin{equation}
\alpha'_{\mu \nu}(B)   =   \alpha  \   \de_{\mu \nu}
\end{equation}
and    write,    for any  $U \in \cB_{j+1}$
 \begin{equation}
 \sum_{\overline{B} =U} V( {_{f}\beta},  \alpha,  B)   =   \sum_{\overline{B} =U} V( {_{f}\beta},  \alpha',  B)    -  \cL_4 ({_{f}K}) (U)   -   \De ({_{f}K}) (U)   
 \end{equation}
 where,   for  $U \subset  \cB_{j+1}$,
 \begin{equation}
 \begin{split}
  \cL_4 ( {_{f}K}) (U)  = &   \cL_4 ({_{0}K}) (U) =\sum_{\overline{B}=U}  V(0,  \alpha' -  \hat \alpha    ,  B)  =   V(0, \alpha'  -  \hat \alpha  , U)  \\
 \De ({_{f}K}) (U)  = &  \De ({_{0}K}) (U) = \sum_{\overline{B}=U}  V(0, \hat \alpha  - \alpha  ,  B) = V(0,  \tilde  \alpha,  U)        \\
 \end{split}
 \end{equation}
 where  $\tilde \alpha_{\mu \nu}  (x)   =   \hat   \alpha_{\mu \nu}  (B)- \alpha_{\mu \nu}  (B) $  if  $x \in B$. Then we can write that $ \cL_4 (\de {_{f}K}) (U) = 0$ and $\De (\de{_{f}K}) (U) =0$

By the above definition  $\De( {_{0}K})(U)$    vanishes  unless  $U$ touches the boundary.     Now     (\ref{SplitL3prime})  becomes
\begin{equation}   \label{grrL3}
\begin{split}
\cL_3 &( {_{f}E}, \sigma, \tilde  {_{f}E}, \tilde  \sigma,  {_{f}K}, {_{0}K})(U) \\
=&   \sum_{ \overline{B}  = U} 
  \left( V(\tilde  {_{f}E}, \tilde  \sigma, B)-V^\#( {_{f}E}, \sigma, B)  -    V( {_{f}\beta},   \alpha',  B)  \right)  \\
& +  \cL'_3 ( {_{0}K}) (U)   +\cL_4 ({_{0}K})(U)  +  \De( {_{0}K})  (U) \\
  \end{split}
  \end{equation}
%\bigskip

{\noindent{\bf Remark}. }Because $\cL_4 ( {_{f}K}) = \cL_4 ( {_{0}K})$ and $\De ({_{f}K}) = \De( {_{0}K})$ are independent from $f $, we will have the same results as Lemma 10 and Lemma 11 in (Dimock, \cite{Dim09}). Moreover, by using Lemma \ref{albeta} above, we can obtain   some explicit upper bounds for $\cL_4 ( {_{0}K})$ and $ \De( {_{0}K})$.

\begin{lem} \label{L4}
 Let  $L$ be sufficiently large.   Then 
the  operator  $\cL_4$  is a  contraction with %arbitrarily small norm
\begin{equation}
\| \cL_4 ({_{0}K})\|_{j+1} \leq 4 (2d)^3 n_2(d) L^{-(j+1)} \| {_{0}K} \|_j
\end{equation}
 \end{lem}
 
\begin{lem}  
 Let  $L$ be sufficiently large.   Then 
the  operator  $\De  $  is a  contraction with %arbitrarily small norm
\begin{equation}
\| \De ({_{0}K}) \| \leq 4 (2d)^5 2^d n_2(d) L^{-1} \| {_{0}K} \|_j
\end{equation}
 \end{lem}

%\bigskip

 \subsection{ Simplifying for the next scale }
 
We  now pick      $\tilde  {_{f}E}(B), \tilde   \sigma$   so  the $V$ terms in (\ref{grrL3}) cancel.  We have:
\begin{equation} \label{114}
\begin{split}
 V^\#( {_{f}E},  \sigma, B, \phi)   
 = & {_{f}E}(B) + \int   \frac {\sigma}{4}  \sum_{x \in B}    \sum_{\mu}  (\partial_{\mu}\phi(x) + \partial_{\mu}  \zeta(x)) ^2
   d \mu_{\Gamma_{j+1}} (\zeta)\\
= & {_{f}E}(B) +     \frac{\sigma}{4}  \sum_{x \in B}  \sum_{\mu}       \partial_{\mu}\phi(x)  ^2   \int    d \mu_{\Gamma_{j+1}} (\zeta)\\
& + \frac {\sigma}{2}  \sum_{x \in B}    \sum_{\mu}  \partial_{\mu}\phi(x) \int \partial_{\mu}  \zeta(x)
   d \mu_{\Gamma_{j+1}} (\zeta)  \\
& +     \frac{\sigma}{4}    \sum_{x \in B}  \sum_{\mu}\int   \partial_{\mu}  \zeta(x) ^2
   d \mu_{\Gamma_{j+1}} (\zeta)\\
= & {_{f}E}(B) +     \frac{\sigma}{4}  \sum_{x \in B}  \sum_{\mu}       \partial_{\mu}\phi(x)  ^2   + 0 +    \frac{\sigma}{4}    \sum_{x \in B}  \sum_{\mu}  ( \partial_{\mu}  \Gamma_{j+1}\partial_{\mu}^*)(x,x)\\
   \equiv  & V({_{f}E}, \sigma, B, \phi)   +   \frac{\sigma}{4}     \sum_{\mu}   Tr ( 1_B   \partial_{\mu} \Gamma_{j+1} \partial^*_{\mu})  \\
\end{split}
\end{equation}
because
\begin{equation}
\begin{split}
\int \partial_{\mu}  \zeta(x)  d \mu_{\Gamma_{j+1}} (\zeta) &= 0 \\
\int   \partial_{\mu}  \zeta(x) ^2    d \mu_{\Gamma_{j+1}} (\zeta) &= \int (\zeta, \partial_{\mu}^* \de_x)  (\zeta, \partial_{\mu}^* \de_x) d \mu_{\Gamma_{j+1}} (\zeta) \\
&= ( \partial_{\mu}^* \de_x, \Gamma_{j+1}  \partial_{\mu}^* \de_x) \\
&= (  \de_x, \partial_{\mu} \Gamma_{j+1}  \partial_{\mu}^* \de_x) \\
&=  ( \partial_{\mu}  \Gamma_{j+1}\partial_{\mu}^*)(x,x)
\end{split}
\end{equation}
If we choose $\tilde {_{f}E}  = \tilde {_{f}E} ( {_{f}E}, \sigma, {_{f}K})$  
\begin{equation} \label{Etilde}
 \tilde    {_{f}E}(B)   =  {_{f}E}(B)     +  \frac{\sigma}{4}     \sum_{\mu}   Tr   ( 1_B   \partial_{\mu} \Gamma_{j+1} \partial^*_{\mu})  
 +   {_{f}\beta}({_{f}K},B)     
\end{equation} 
then  the constant terms of (\ref{114}) will be canceled. The   second  order    terms  of  (\ref{114}) would be vanish  if we  define  $\tilde \sigma = \tilde \sigma (\sigma,{_{f}K}) = \tilde \sigma (\sigma,{_{0}K})$ by
\begin{equation}   
  \tilde   \sigma =      \sigma     +   \alpha  ({_{f}K})   =    \sigma     +   \alpha  ({_{0}K})
\end{equation}
Here we are canceling the constant term  exactly for all $B$,   but  for  the quadratic term  we   only cancel  the  invariant  version  away from the boundary.

By  composing   $ {_{f}K}' =  {_{f}K}'( \tilde{_{f}E}, \tilde   \sigma, {_{f}E}, \sigma, {_{f}K}, {_{0}K})$ in theorem \ref{T5} with newly defined  $\tilde {_{f}E} =\tilde  {_{f}E} ( {_{f}E}, \sigma,  {_{f}K})$  
  and   $\tilde \sigma = \tilde \sigma (\sigma,{_{f}K}) = \tilde \sigma (\sigma,{_{0}K})$ 
we obtain  a new map    ${_{f}K}' = {_{f}K}'( {_{f}E}, \sigma, {_{f}K}, {_{0}K})$.     We  also have new quantities  
$ {_{f}E}'({_{f}E}, \sigma,  {_{f}K})$  defined by  
 ${_{f}E}'(B' )  =  \sum_{B \subset B'}\tilde  {_{f}E}(B)$  and  $\sigma'  =  \sigma'(\sigma, {_{f}K}) =  \sigma'(\sigma, {_{0}K})$   defined by  $\sigma'=  \tilde \sigma   =   \sigma     +    \alpha({_{f}K}) =  \sigma     +    \alpha({_{0}K})$ as normal.  These quantities satisfy    (\ref{RG})
\begin{equation}  \label{rg118}
 \mu_{\Gamma_{j+1}} * \left(  {_{f}I} ({_{f}E},\sigma)     \circ {_{f}K}  \right)(\Lambda) =  \left( {_{f}I}'( {_{f}E}',\sigma')     \circ  {_{f}K}'  \right)(\Lambda) 
 \end{equation}

 Here we  still assume that  $L$ is sufficiently large, and that  $A$ is sufficiently large depending on $L$.

  \begin{thm} \label{T6}
 \begin{enumerate}
 \item For        $R>0$    there   is a $ r>0$  such that   the following holds  for all j.    If    $  \| {_{f}E}\|_j $,  $| \sigma|$, $\max  \{  \| {_{f}K} \|_j ,  \| {_{0}K} \|_j  \} < r$     then   $ \| {_{f}E}'\|_{j+1}$, $|\sigma' |$,  $\max  \{ \| {_{f}K}'\|_{j+1}, \| {_{0}K}'\|_{j+1} \}  <  R$.  Furthermore   $ {_{f}E}', {_{f}K}', \sigma' $   are   smooth functions  of   $ {_{f}E}, \sigma, {_{f}K}, {_{0}K}$    on this domain with derivatives bounded uniformly in $j$. The analyticity of ${_{f}K}'$ in $t_1,\dots,t_m$ still holds when we go from $j$-scale to $(j+1)$-scale.
 \item   
 The      linearization   of  ${_{f}K}' = {_{f}K}'({_{f}E}, \sigma, {_{f}K}, {_{0}K})$  at the origin is the   contraction  $\cL( {_{f}K})$ where
 \begin{equation}
 \cL=  \cL_1  + \cL_2 + \cL'_3 +  \cL_4  + \De
\end{equation}
\end{enumerate}
\end{thm}

{\noindent{\bf Proof}. }   For the first  part, by combining with theorem \ref{T5}, it suffices to show that the linear maps  $\tilde {_{f}E}$  and   $\tilde \sigma$    have norms  bounded uniformly  in   $j$. 
Using the estimate   $|\alpha( {_{f}K})|   =|\alpha( {_{0}K})|   \leq 4 (2d)^2 n_2(d) h^{-2} A^{-1} \| {_{0}K}\|_j$ from  lemma \ref{albeta}, we have $\tilde \sigma$   is bounded.  From  lemma \ref{albeta} we also have  the bound on $\| {_{f}\beta}( {_{f}K})\|_j  \leq  2n_2(d) A^{-1}  \| {_{f}K} \|_j$. For   $B \in \cB_j$, the estimate  (\ref{imp})   gives  us
\begin{equation}
\left|  \frac{\sigma}{4}     \sum_{\mu}   Tr   ( 1_B   (\partial_{\mu} \Gamma_{j+1} \partial^*_{\mu})  \right|
\leq d c_{1,1}   |\sigma|   \sum_{x \in B}  L^{-dj}  \leq   d c_{1,1}   |\sigma|  
\end{equation}
where $c_{1,1}$ as in (\ref{imp}).
Combining with (\ref{Etilde}) we  have  that $\tilde {_{f}E}  = \tilde {_{f}E}( {_{f}E}, \sigma, {_{f}K})$ satisfies
\begin{equation}  \label{121}
\| \tilde  {_{f}E} \|_j   \leq   \| {_{f}E}\|_j  +    \cC ( |\sigma|   +  A^{-1}  \| {_{f}K} \|_j  )
\end{equation}
where $\cC = \max \{ d c_{1,1}, 2 n_2 (d)$. 

The  second part  follows  since the linearization  of the new function ${_{f}K}'$  is the linearization 
of the old function  ${_{f}K}'$ in theorem \ref{T5} composed  with $\tilde {_{f}E} =\tilde  {_{f}E}( {_{f}E}, \sigma,  {_{f}K}),   \tilde \sigma   = \tilde  \sigma(\sigma , {_{f}K}) = \tilde  \sigma(\sigma , {_{0}K})$.
(All of them vanish at zero.)   The cancellation  gives us only with  $\cL( {_{f}K})$.

  \subsection{ Forming RG FLow}

 It is easier for us if we can extract the energy from the other variables.   Assume that we start with $E(B)=0$ in (\ref{rg118})  
\begin{equation} 
 \mu_{\Gamma_{j+1}} * \left( {_{f}I}(0,\sigma)   \circ  {_{f}K}  \right)(\Lambda_N) 
 =  \left(  {_{f}I}'( {_{f}E}',\sigma')  \circ  {_{f}K}' \right)(\Lambda) \\
 \end{equation}
where            $\sigma'  = \sigma'( \sigma, {_{f}K}) = \sigma'( \sigma, {_{0}K})$  and  ${_{f}K}' = {_{f}K}'(0, \sigma, {_{f}K})$
and    ${_{f}E}'  = {_{f}E}'(0, \sigma, {_{f}K}) $ as above. Then we remove the  $ {_{f}E}'$ by  making an adjustment  in ${_{f}K}'$. 
\begin{equation} \label{123}
\begin{split}
& \mu_{\Gamma_{j+1}} * \left( {_{f}I}(0,\sigma)   \circ  {_{f}K}  \right)(\Lambda_N) 
=  \left(  {_{f}I}'( {_{f}E}',\sigma')  \circ  {_{f}K}' \right)(\Lambda) \\
&= \sum_{U \in \cP_{j+1}}\left( {_{f}I}'( {_{f}E}',\sigma') (\Lambda -  U) \right)  \left({_{f}K}'(0, \sigma, {_{f}K}, U)   \right) \\
&=\sum_{U \in \cP_{j+1}} \left(  \prod_{ B' \in \cB_{j+1} (\Lambda - U) }   {_{f}I}'( {_{f}E}',\sigma') (B')    \right)   \left({_{f}K}'(0, \sigma, {_{f}K}, U)   \right)          \\
&= \sum_{U \in \cP_{j+1}} \left(  \prod_{ B' \in \cB_{j+1} (\Lambda - U) } \exp ( {_{f}E}'(B')) [ {_{f}I}'(0,\sigma') (B') ]   \right)   \left({_{f}K}'(0, \sigma, {_{f}K}, U)   \right)\\         
&= \sum_{U \in \cP_{j+1}} \left(   \exp \left( \sum_{ B' \in \cB_{j+1} (\Lambda - U)} {_{f}E}'(B') \right)   {_{f}I}'(0,\sigma') (\Lambda - U)  \right)   \left({_{f}K}'(0, \sigma, {_{f}K}, U)   \right)\\
&= \exp \left(   \sum_{B'  \in \cB_{j+1}(\Lambda_N)}  {_{f}E}^+(B') \right) \left(  {_{f}I}'(0,\sigma^+)     \circ  {_{f}K}^+ \right)(\Lambda_N)\\
\end{split}
\end{equation}
where ${_{f}E}^+(\sigma, {_{f}K}, B'),  \sigma^+(\sigma , {_{f}K}),  {_{f}K}^+(\sigma, {_{f}K}, U)$ are defined as following  ($U \in \cP_{j+1}, B'  \in  \cB_{j+1}$)
\begin{equation}   \label{a}
\begin{split}   
{_{f}E}^+(\sigma, {_{f}K}, B') \equiv &  {_{f}E}'(0, \sigma, {_{f}K}, B')  =  \sum_{B \subset  B'}  \tilde  {_{f}E}( 0, \sigma, {_{f}K}, B)    \\
    \sigma^+(\sigma , {_{f}K})  \equiv &  \sigma'(\sigma , {_{f}K})  =   \sigma'(\sigma , {_{0}K}) =    \sigma+   \alpha({_{0}K})     \\
    {_{f}K}^+(\sigma, {_{f}K}, U)  \equiv    &\exp \left(  - \sum_{ B' \in \cB_{j+1}(U) } {_{f}E}^+(B')   \right) {_{f}K}'(0,\sigma, {_{f}K},  U)  \\
\end{split}
\end{equation}

The dynamical  variables are  now     $ \sigma^+(\sigma , {_{f}K})$    and   $ {_{f}K}^+(\sigma,{_{f}K})$.   The extracted  energy       $ {_{f}E}^+(\sigma,  K)$  is controlled by the other variables. Because  everything vanishes at the origin    the linearization of  ${_{f}K}^+(\sigma, {_{f}K})$  is  still  $\cL( {_{f}K)}$.  The bound  (\ref{121})  on $\tilde {_{f}E}$  would  give us an upper bound  on  ${_{f}E}^+$ and   our   theorem \ref{T6} becomes:
  \begin{thm}  \label{T7}
 \begin{enumerate}
 \item For       $R>0$    there   is a $r >0$  such that the following holds  for all j.    If     $| \sigma|$, $\max   \{ \| {_{f}K} \|_j ,   \| {_{0}K} \|_j   \} <   r$   then   $|\sigma^+ |$,  $\max   \{ \| {_{f}K}^+\|_{j+1} , \| {_{0}K}^+\|_{j+1} \} <  R$.  Furthermore   $ \sigma^+, {_{f}K}^+ $   are   smooth functions     of   $ \sigma, {_{f}K}$    on this domain with derivatives bounded uniformly in $j$. The analyticity of ${_{f}K}^+$ in $t_1,\dots,t_m$ still holds when we go from $j$-scale to $(j+1)$-scale.
 \item   The extracted energies satisfy 
 \begin{equation}  \label{125}
\| {_{f}E}^+( \sigma, {_{f}K})\|_{j+1}  \leq   \cC(L^d) \Big(  |\sigma|   + A^{-1}  \|  {_{f}K} \|_j  \Big)
 \end{equation}
 \item   
 The      linearization   of  $K^+$  at the origin is the contraction   $\cL$.
\end{enumerate}
\end{thm}

\section{The  stable manifold}  \label{stable}
Up to now, we   have not   specialized to the dipole  gas,  but  take  a  general  initial    point   $\sigma_0,  {_{f}K}_0$   corresponding to an    integral     $\int  ({_{f}I}(0, \sigma_0)\circ  {_{f}K}_0) (\Lambda_N)  d \mu_{C_0}$. 
We assume   ${_{0}K}_0(X, \phi)$ has the lattice symmetries      and satisfies   the conditions  (\ref{symmetry}). We  also  assume   $|\sigma_0|, \max \{ \| {_{f}K} \|_0 , \| {_{0}K} \|_0 \}  < r $  where   $r$   is small enough so the  theorem \ref{T7} holds,  say with $R =1$,  then we can take the first step. We  apply the transformation (\ref{123}) for  $j=0,1,2, \dots$   and continue as far as we can. Then we get a sequence   $\sigma_j, {_{f}K}^N_j(X)$    by  $\sigma_{j+1}  =  \sigma^+(\sigma_j,  {_{f}K}^N_j)$    and   ${_{f}K}^N_{j+1}  = {_{f}K}^+ (  \sigma_j, {_{f}K}^N_j)$
 with    extracted energies      $ {_{f}E}^N_{j+1}   =    {_{f}E}^+(\sigma_j,  {_{f}K}^N_j) $. Then  we  have, for any $l$, with  ${_{f}I}_j(\sigma_j)  = {_{f}I}_j(0, \sigma_j)$ 
 \begin{equation}  \label{ksteps}
\begin{split}
\int   ({_{f}I}_0(\sigma_0) \circ  {_{f}K}_0 )(\Lambda_N)  d \mu_{C_0} =    \exp \left(  \sum_{j=1}^l \sum_{B \in \cB_j(\Lambda_N)}     {_{f}E}^N_j(B)   \right)
 \int   ({_{f}I}_l(\sigma_l) \circ  {_{f}K}^N_l )(\Lambda_N) d \mu_{C_l}
\end{split}
\end{equation}
 The quantities      ${_{0}K}^N_j(X)$   and  ${_{0}E}^N_j(B)$  are independent of  $N$ and   have  the lattice symmetries   if   $X,B$  are away from  the boundary $\partial  \Lambda_N$ in the sense that   they have no boundary  blocks.     These properties  are true initially  and  are preserved by the   iteration.    In these cases  we denote these  quantities by  just    ${_{0}K}_j(X)$   and  ${_{0}E}_j(B)$

With  our  construction  $\alpha$ defined  in (\ref{DefAlBe}),(\ref{defsymalpha})  only depends on   ${_{0}K}_j$. By  splitting  $K^+$  into a  linear  and  a higher order piece  the sequence   $\sigma_j, {_{f}K}^N_j(X)$    is generated by   the RG transformation
\begin{equation}  \label{flow1}
\begin{split}
\sigma_{j+1} = &  \sigma_j + \alpha(K_j)  \\
{_{0}K}^N_{j+1}  =&  \cL({_{0}K}^N_j)   +  {_{0}g}(\sigma_j, {_{0}K}^N_j)\\
\de{_{f}K}^N_{j+1}  =& \left( \cL_1 + \cL_2 \right) (\de{_{f}K}^N_j)   +  {_{f}g}(\sigma_j, {_{f}K}^N_j, {_{0}K}^N_j) -  {_{0}g}(\sigma_j, {_{0}K}^N_j)\\
\end{split}
\end{equation}
This is regarded as a mapping from the Banach space  ${\mathbb{R}}  \times  \left( {{\cal K}}_j(\Lambda_N) \times {{\cal K}}_j(\Lambda_N) \right)$ to the Banach space    ${\mathbb{R}}  \times  \left( {{\cal K}}_j(\Lambda_N) \times {{\cal K}}_j(\Lambda_N) \right)$. The 2nd equation of (\ref{flow1}) defines ${_{0}g}$ which is smooth  with  derivatives bounded  uniformly   in  $j$  and satisfies   ${_{0}g}(0,0) =0$,  $D({_{0}g})(0,0)=0$. The last  equation of (\ref{flow1}) defines ${_{f}g}$ which is also   smooth  with  derivatives bounded  uniformly   in  $j$  and satisfies   ${_{f}g}(0,0) =0$,  $D({_{f}g})(0,0)=0$.

Now we consider  the first  two equations in  (\ref{flow1}).  Around the origin there   are  a neutral  direction  $\sigma_j$   and  a contracting direction  $K_j$  (since $\cal{L}$ is    a  contraction.).    Hence we expect there is a stable manifold.   We  quote a version of  the stable manifold theorem due to Brydges \cite{Bry07},    as applied   in  Theorem 7 in Dimock \cite{Dim09}

\begin{thm}   \label{stable}  {(Theorem 7, Dimock \cite{Dim09})}

Let  $L$ be sufficiently large,  $A$ sufficiently large  (depending on $L$), and  $r$ sufficiently small (depending on $L,A$).  Then there is     $0 < \tau <  r$  and     a smooth  real-valued   function       $  \sigma_0  = h({_{0}K}_0),\  h(0) =0$,  mapping          $\|  {_{0}K}_0 \|_0  < \tau$   into  $|\sigma_0| < r$  such that with these start values   the sequence 
$\sigma_j, {_{0}K}^N_j$    is   defined for all  $0 \leq j \leq  N$  and    
\begin{equation} \label{kingKs}
|\sigma_j|   \leq  r 2^{-j}   { \hspace{1cm}}   \|{_{0}K}^N_j \|_j  \leq     r 2^{-j} 
\end{equation}
Furthermore  the  extracted    energies satisfy  
\begin{equation}  \label{energy}
\|  {_{0}E}^N_{j+1}  \|_{j+1}  \leq   2 \cC( L^d)   r 2^{-j} 
\end{equation}
\end{thm}

%\bigskip

{\noindent{\bf Remark}. }Using the Lemma \ref{initialbounds} below, given $r>0$, we can always choose $z, \sigma_0$ and $\max_k |t_k|$ sufficiently small then 
$\max \{ \| {_{f}K} \|_0 , \| {_{0}K} \|_0 \}  \leq r$. 
Now we claim that $ \| {_{f}K}^N_j \|_j$ has the same  bound as the $ \|{_{0}K}^N_j \|_j $ in the last theorem.

Supposed that at $j=k$, we have: $ \| {_{f}K}^N_j \|_j  \leq r 2^{-k}$ . As in the proof of Theorem 7 in (Dimock, \cite{Dim09}),  we can say that $\cL$ and $(\cL_1 + \cL_2)$ is a contraction with  norm less  than  $1 /8$  and   $ {_{f}g}(\sigma_j, {_{f}K}^N_j, {_{0}K}_j)$  is second order.  Hence there are some constant $H$ such that:
 $\|  {_{f}g}(\sigma_j, {_{f}K}^N_j, {_{0}K}^N_j) \| \leq H  \Big(   |\sigma_j|^2  +   \|{_{0}K}^N_j \|_j^2 +  \| {_{f}K}^N_j \|_j^2\Big)$ with $|\sigma_j|,  \|{_{0}K}^N_j \|_j,  \|{_{f}K}^N_j \|_j$ small.
 Then we have:
\begin{equation}
\begin{split}
\|  {_{f}K}^N_{j+1}  \|_{j+1}     \leq  & \frac{1}{8} \left( \|  {_{0}K}^N_j \|_j +  \| \de {_{f}K}^N_j \|_j    \right)  +  H \Big(   |\sigma_j|^2  +   \|{_{0}K}^N_j \|_j^2 +  \|{_{f}K}^N_j \|_j^2   \Big) \\
 \leq  & \frac{1}{8} \left( 2 \|  {_{0}K}^N_j \|_j +  \| {_{f}K}^N_j \|_j    \right)  +   3H   \Big(     r 2^{-j}    \Big)^2 \\
\leq  &\ \frac{1}{8}   \Big(  3   r 2^{-j}  \Big) +   3H   \Big(     r 2^{-j}    \Big)^2  \\
\leq  &      r 2^{-j-1} \\
\end{split}
\end{equation}
for $r$ sufficiently small

The bound for  $\| {_{f}E}^N\|_{j+1}$ comes from the bound on $\sigma_j,  \|  {_{f}K}^N_j \|_j $ , (\ref{125}) and $A>1$.
%\noindent 

Combining with the last theorem, for all  $0 \leq j \leq  N$  we can have:    
\begin{equation} \label{kingKs}
|\sigma_j|   \leq  r 2^{-j}   { \hspace{1cm}}   \|{_{f}K}^N_j \|_j  \leq     r 2^{-j} 
\end{equation}
and  the  extracted    energies satisfy  
\begin{equation}  \label{energy}
\|  {_{f}E}^N_{j+1}  \|_{j+1}  \leq   2 \cC( L^d)   r 2^{-j}. 
\end{equation}

\section{The dipole gas}    \label{dipole}
\subsection{The initial  density}

Now we consider the  generating function:
\begin{equation}
{_{f}Z}_N(z,\sigma)=  \int  e^{i f(\phi)} \exp  \Big(zW( \Lambda_N,   \sqrt{1+\sigma} \phi)- \sigma  V( \Lambda_N, \phi)\Big)     d\mu_{C_0} (\phi)
\end{equation}
When $f=0$, it becomes
\begin{equation}
{_{0}Z}_N(z,\sigma)=  \int  \exp  \Big(zW( \Lambda_N,   \sqrt{1+\sigma} \phi)- \sigma  V( \Lambda_N, \phi)\Big)     d\mu_{C_0} (\phi)
\end{equation}
For  $B \in \cB_0$, we define: $W_0(B)  =  z W(\sqrt{1+ \sigma_0},  B)$  as  in    (\ref{defWuB}) and  $V_0(B)  =  \sigma_0V( B)$  as  in      (\ref{41}). Then we follow with   a Mayer expansion to   put the density in the form we want.
\begin{equation}  \label{sink}
\begin{split} 
{_{f} \cZ}^N_0   =&    \prod_{B \subset  \Lambda_N} e^{ i f(\phi) +  W_0(B)-V_0(B)} \\
 =&    \prod_{B \subset  \Lambda_N} \Big(e^{ - V_0(B) } + (e^{ i f(\phi) +  W_0(B)}-1)e^{ - V_0(B)}  \Big) \\
=&   \sum_{X \subset   \Lambda_N } {_{f}I}_0(\sigma_0,\Lambda_N - X)  {_{f}K}_0(X) \\
=&  (   {_{f}I}_0(\sigma_0) \circ  {_{f}K}_0 )(\Lambda_N)\\
\end{split}
\end{equation}
where  $ I_0(\sigma_0,B)=  e^{ - V_0(B)}$   and   ${_{f}K}_0(X)   = {_{f}K}_0( z,  \sigma_0, X)$   is  given by  
\begin{equation}
{_{f}K}_0( X )  = \prod_{B  \subset  X}  (e^{  if(\phi)|_{B} + W_0(B)}-1)e^{- V_0( B)}
\end{equation}
when $f(\phi) = \sum_{k=1}^m t_k \exp \left( i\partial_{\mu_k} \phi(x_k) \right)$,  $ if(\phi)\big|_{B}=  t_k \exp \left( i\partial_{\mu_k} \phi(x_k) \right)$ if $B =\{x_k\}$ for some $k$, otherwise $ if(\phi)\big|_{B}=0$

or 
\begin{equation}
{_{f}K}_0( X )  = \prod_{B  \subset  X}  (e^{ if(\phi)|_{B} + W_0(B)}-1)e^{- V_0( B)}
\end{equation}
when $f(\phi) =f(\phi) = \sum_{k=1}^m t_k \partial_{\mu_k} \phi (x_k)$, $ if(\phi)\big|_{B}=  t_k \partial_{\mu_k} \phi (x_k)$  if $B =\{x_k\}$ for some $k$, otherwise $ if(\phi)\big|_{B}=0$

or 
\begin{equation}
{_{f}K}_0( X )  = \prod_{B  \subset  X}  (e^{W_0(B)}-1)e^{- V_0( B)}
\end{equation}
when $f(\phi) = 0$

Note that, when $f = 0$ ,    ${_{0}K}_0$ actually is the $K_0$ in lemma 12,  \cite{Dim09}. We also can prove the same result for ${_{f}K}_0$.
\begin{lem}  \label{initialbounds}
{ \ }
Given  $1>r>0$ , there are some sufficiently small $a(r), b(r)$ and $c(r)$ such that if $\max_k |t_k| \leq a(r)$,  $|z| \leq b(r)$ and $|\sigma_0| \leq c(r)$  then   $\|  {_{f}K}_0(z,  \sigma_0)  \|_0  \leq  r$.
Furthermore   ${_{f}K}_0$ is a smooth function of  $(z,  \sigma_0)$, and analytic in $t_k$ for all $k=1,\dots, m$.
\end{lem}

{\noindent{\bf Proof}. }

*When $f=0$,  using lemma 12 \cite{Dim09}, we have some $b_0(r), c_0(r)$ such that  $\|  {_{0}K}_0(z,  \sigma_0)  \|_0  \leq  r$ if $|z| \leq b_0(r)$ and $|\sigma_0| \leq c_0(r)$ 

*In the case  $f(\phi) = \sum_{k=1}^m t_k \partial_{\mu_k} \phi (x_k)$, using (\cite{Dim09}, (95)), for $\phi = \phi' + \zeta$, we have: 
\begin{equation}
\begin{split}
&\|  (e^{   if(\phi)|_{B} + W_0(B)}-1)\|_0 = \|  (e^{   if(\phi)|_{B} + z W(\sqrt{1+ \sigma_0},  B)}-1)\|_0\\
&\leq \sum_{n=1}^\infty  \frac{1}{n!} \|   z W(\sqrt{1+ \sigma_0},  B) +     if(\phi)|_{B}   \|_0^n    \\
&\leq \sum_{n=1}^\infty  \frac{1}{n!} \left( \|   z W(\sqrt{1+ \sigma_0},  B) \|_0 +  \|   if(\phi)|_{B}   \|_0 \right)^n   \\
&\leq \sum_{n=1}^\infty  \frac{1}{n!} \left(2  |z|  e^{h\sqrt{d(1+\sigma_0)}} +  \max_k |t_k|h^{-1}\| \phi  \|_{\Phi_0(B^*)} \right)^n   \\
\end{split}
\end{equation}

We can assume that $\max_k |t_k|h^{-1} \leq 1$. Applying  lemma  3 in \cite{Dim09}, we get    $\|  e^{-V_0(B ) }\|_{s,0}        \leq   2$.
\begin{equation}
\begin{split}
&\| {_{f}K}_0(B) \|_0     =      \sup_{ \phi', \zeta } \|  {_{f}K}_0(B, \phi' + \zeta) \| _0     G_0(X, \phi', \zeta)^{-1}\\
&\leq  \sup_{ \phi', \zeta } \|  (e^{   if(\phi)|_{B} + W_0(B)}-1) \|_0  \| e^{- V_0( B)} \| _0     G_{s,0}(X, \phi', \zeta)^{-2}\\
&\leq \| e^{- V_0( B)} \| _{s,0}  \sup_{ \phi', \zeta } \|  (e^{   if(\phi)|_{B} + W_0(B)}-1) \|_0  G_{s,0}(X, \phi', \zeta)^{-1}\\
&\leq2  \sup_{ \phi', \zeta } \left(  \exp \left(2  |z|  e^{h\sqrt{d(1+\sigma_0)}} +  \max_k |t_k|h^{-1} \| \phi'+ \zeta  \|_{\Phi_0(B^*)}  \right) -1  \right)  G_{s,0}(X, \phi', \zeta)^{-1}\\
&\leq2  \sup_{ \phi', \zeta } \left(  \exp \left(2  |z|  e^{h\sqrt{d(1+\sigma_0)}} \right) -1\right) \exp \left(   \max_k |t_k|h^{-1} \| \phi'+ \zeta  \|_{\Phi_0(B^*)} \right)G_{s,0}(X, \phi', \zeta)^{-1}\\
&+ 2  \sup_{ \phi', \zeta } \left(  \exp  \left( \max_k |t_k|h^{-1} \| \phi'+ \zeta  \|_{\Phi_0(B^*)} \right) -1  \right)
 G_{s,0}(X, \phi', \zeta)^{-1}  \\
%\end{split}
%\end{equation}
%\begin{equation}
%\begin{split}
&\leq2  \sup_{ \phi', \zeta } \left(  \exp \left(2  |z|  e^{h\sqrt{d(1+\sigma_0)}} \right) -1\right) \exp \left(   \| \phi'+ \zeta  \|_{\Phi_0(B^*)} \right) e^{- \| \phi'  \|_{\Phi_0(B^*)}^2 - \| \zeta  \|_{\Phi_0(B^*)}^2 }\\
&+ 2  \sup_{ \phi', \zeta } \left(  \exp  \left( \max_k |t_k|h^{-1} \| \phi'+ \zeta  \|_{\Phi_0(B^*)} \right) -1  \right)  e^{- \| \phi'  \|_{\Phi_0(B^*)}^2 - \| \zeta  \|_{\Phi_0(B^*)}^2}  \\
\end{split}
\end{equation}
Because $\exp \left(   \| \phi'+ \zeta  \|_{\Phi_0(B^*)} \right) \exp (- \| \phi'  \|_{\Phi_0(B^*)}^2 - \| \zeta  \|_{\Phi_0(B^*)}^2 )$ is bounded and  
\begin{equation}
\lim_{z,\sigma_0  \rightarrow0} \left( \exp \left(2  |z|  e^{h\sqrt{d(1+\sigma_0)}} \right) -1 \right) =0 
\end{equation}
there exist some sufficiently small $b_1(r), c_1(r) > 0$ such that we have
\begin{equation}
\begin{split}
2  \sup_{ \phi', \zeta } \left(  \exp \left(2  |z|  e^{h\sqrt{d(1+\sigma_0)}} \right) -1\right) \exp \left(   \| \phi'+ \zeta  \|_{\Phi_0(B^*)} \right) e^{- \| \phi'  \|_{\Phi_0(B^*)}^2 - \| \zeta  \|_{\Phi_0(B^*)}^2 }  \leq \frac{r}{4A}
\end{split}
\end{equation}
for all $|z| \leq b_1(r)$ and $|\sigma_0| \leq c_1(r)$.

 For other part, we have:
\begin{equation}
\begin{split}
&2  \sup_{ \phi', \zeta } \left(  \exp  \left( \max_k |t_k|h^{-1} \| \phi'+ \zeta  \|_{\Phi_0(B^*)} \right) -1  \right)  \exp (- \| \phi'  \|_{\Phi_0(B^*)}^2 - \| \zeta  \|_{\Phi_0(B^*)}^2 )\\
& \leq 2  \sup_{ \phi', \zeta } \left(  \exp  \left( \| \phi'   \|_{\Phi_0(B^*)} + \|\zeta  \|_{\Phi_0(B^*)} \right) -1  \right)  \exp (- \| \phi'  \|_{\Phi_0(B^*)}^2 - \| \zeta  \|_{\Phi_0(B^*)}^2 ) \\
\end{split}
\end{equation}

We also can find some sufficiently large $H$ such that if $ \| \phi'   \|_{\Phi_0(B^*)} + \|\zeta  \|_{\Phi_0(B^*)} \geq H$ then
\begin{equation}
 2\left(  \exp  \left( \| \phi'   \|_{\Phi_0(B^*)} + \|\zeta  \|_{\Phi_0(B^*)} \right) -1  \right)  \exp (- \| \phi'  \|_{\Phi_0(B^*)}^2 - \| \zeta  \|_{\Phi_0(B^*)}^2 ) \leq \frac{r}{4A}
\end{equation}

For  $ \| \phi'   \|_{\Phi_0(B^*)} + \|\zeta  \|_{\Phi_0(B^*)} \leq H$, we have $\| \phi'  + \zeta  \|_{\Phi_0(B^*)} \leq \| \phi'   \|_{\Phi_0(B^*)} + \|\zeta  \|_{\Phi_0(B^*)}\leq H$. 
So with  $\max_k |t_k| \leq a_1(r)$ sufficiently small and  $ \| \phi'   \|_{\Phi_0(B^*)} + \|\zeta  \|_{\Phi_0(B^*)} \leq H$, 
\begin{equation}
 2\left(  \exp  \left( \max_k |t_k|h^{-1} \| \phi'+ \zeta  \|_{\Phi_0(B^*)} \right) -1  \right)  \exp (- \| \phi'  \|_{\Phi_0(B^*)}^2 - \| \zeta  \|_{\Phi_0(B^*)}^2 ) \leq  \frac{r}{4A}
\end{equation}

In summary we can always choose sufficiently small $a(r), b(r), c(r)$ such that if  $\max_k |t_k| \leq a_1(r)$, $|z| \leq b_1(r)$, and $|\sigma_0| \leq c_1(r)$ then
\begin{equation}
\| {_{f}K}_0(B) \|_0 \leq 2\frac{r}{4A} = \frac{r}{2A} { \hspace{1cm}} \forall B\in \cB_0
\end{equation}
For those $a_1(r), b_1(r), c_1(r)$, $\max_k |t_k| \leq a_1(r)$, $|z| \leq b_1(r)$, and $|\sigma_0| \leq c_1(r)$  we have
\begin{equation}
\begin{split}
\| {_{f}K}_0 \|_0 &= \sup_{X \in \cP_{0,c}} \| {_{f}K}_0(X) \|_0 A^{|X|_0}  \\
&\leq  \sup_{X \in \cP_{0,c}}  \left(  \prod_{B \subset X} \| {_{f}K}_0(B) \|_0\right)   A^{|X|_0}\\
&\leq  \sup_{X \in \cP_{0,c}}  \left(   \frac{r}{2A} \right)^{|X|_0}   A^{|X|_0} \leq  \frac{r}{2} < r\\
\end{split}
\end{equation}

* In the last case, $f(\phi) = \sum_{k=1}^m t_k \exp \left( i\partial_{\mu_k} \phi(x_k) \right)$, we have:
\begin{equation}
\begin{split}
\|  (e^{  if( \phi)|_B + W_0(B)}-1)\|_0 &\leq \sum_{n=1}^\infty  \frac{1}{n!} \left(2  |z|  e^{h\sqrt{d(1+\sigma_0)}} + \max_k |t_k| \right)^n   \\
&= \exp \left(2 |z|  e^{h\sqrt{d(1+\sigma_0)}} + \max_k |t_k| \right) -1
\end{split}
\end{equation}
Using the same argument as above, we can choose some sufficiently small $a_2(r)$, $b_2(r)$, $c_2(r)$ such that  $\|  {_{f}K}_0(z,  \sigma_0)  \|_0  \leq  r$ when $\max_k |t_k| \leq a_2(r)$,  $|z| \leq b_2(r)$ and $|\sigma_0| \leq c_(r)$. 

Now we just simply pick $a(r) = \max \{ a_1(r), a_2(r)\}$, $b(r) = \max \{ b_0(r), b_1(r), b_2(r)\}$ and $c(r) = \max \{ c_0(r), c_1(r), c_2(r)\}$.

The smoothness follows similarly from Lemma 12, (Dimock, \cite{Dim09}).
\footnote{Instead of using the usual estimates, such as $\left(1 + \| \phi \|_{\Phi_j (B^*)}^2 \right) \leq \exp \left(   \| \phi \|_{\Phi_j (B^*)}^2 \right) = G_{s,j} (B, \phi, 0)$, we can use 
\begin{equation}
\left( 1 + \| \phi \|_{\Phi_j (B^*)}^2 \right) = k \left( \frac{1}{k} + \frac{1}{k} \| \phi \|_{\Phi_j (B^*)}^2  \right)     \leq k \exp  \left(  \frac{1}{k}  \| \phi \|_{\Phi_j (B^*)}^2   \right) = k G_{s,j}^{\frac{1}{k}} (B, \phi, 0)
\end{equation} for any positive integer $k$, and so forth}

{\noindent{\bf Remark}. }We have ${_{f}K}_0$ is analytic. For each step when we jump from $j$-scale to $(j+1)$-scale, the analyticity of ${_{f}K}$ still holds for the next scale.

\bigskip

Noticing that ${_{0}K}_0$ is just the $K_0$ in Section 6, (Dimock, \cite{Dim09}), we need the following lemma to apply Theorem \ref{stable}.
 \begin{lem}  {(Lemma 13, \cite{Dim09})}

 The equation   $\sigma  =   h(  {_{0}K}_0(z, \sigma ))$  defines   a  smooth  implicit function  $\sigma  = \sigma (z)$
 near the origin which satisfies  $\sigma(0) =0$.
 \end{lem}

 Taking  $|z|$  sufficiently small and choosing    $\sigma_0 = \sigma(z)$ ,  we  can apply  theorem \ref{stable}.    For  $0 \leq  l \leq  N$, we  have  
 \begin{equation}    \label{afterflow}
 \begin{split}
 {_{f}Z}_N   = &   \exp \left(  \sum_{j=1}^l \sum_{B \in \cB_j(\Lambda_N)}     {_{f}E}^N_j( B)   \right)
\int   ({_{f}I}_l(\sigma_l) \circ  {_{f}K}^N_l )(\Lambda_N) d   \mu_{C_l}\\
\end{split}  
\end{equation}
where   $|\sigma_j|  \leq  r2^{-j}$  and  $ \| {_{f}K}^N_j\|_j  \leq  r2^{-j}$  
and    $\|{_{f}E}^N_{j+1}\|_{j+1}  \leq  {{\cal O}}(L^d)  r 2^{-j}$.

\subsection{Completing    the proof of Theorem  \ref{T1}}

\begin{thm}  \label{six}
For    $|z|$ and $\max_k |t_k|$ sufficiently small    the following  limit  exists:
\begin{equation}
  \lim_{N\to \infty}  |\Lambda_N|^{-1}  \log  {_{f}Z} _N'(z,  \sigma(z))   
  \end{equation} 
 \end{thm}
 \bigskip
 
{\noindent{\bf Proof}. }

With updated index, the proof can go exactly the same as the proof of Theorem 8, \cite{Dim09}. 
We take      $l=N$  in   (\ref{afterflow}).    At this scale there is only one     block  $\Lambda_N \in \cB_N(\Lambda_N)$ and so   we have
\begin{equation}  \label{152}
\begin{split}
  |\Lambda_N|^{-1}  \log  {_{f}Z}'_N(z, \sigma(z))   =&   |\Lambda_N|^{-1}  \sum_{j=1}^{N} \sum_{B \in \cB_j(\Lambda_N)}     {_{f}E}^N_j(B)   \\
     + &   | \Lambda_N|^{-1} \log \left(   \int \left[  {_{f}I}_N(\sigma_N, \Lambda_N)   +   {_{f}K}^N_N (\Lambda_N) \right]d \mu_{ C_N}  \right)\\
\end{split}     
\end{equation}
The second  term has the form 
\begin{equation}  \label{153}
  | \Lambda_N|^{-1} \log  \left(  1   +  \int       {_{f}F}_N d \mu_{ C_N}  \right)
\end{equation}
where
\begin{equation}
\begin{split} \label{tail}
{_{f}T}_N =&\left(  1   +  \int       {_{f}F}_N d \mu_{ C_N}  \right)\\
{_{f}F}_N(\Lambda_N) = {_{f}F}_N =&    {_{f}I}_N(\sigma_N, \Lambda_N)  -1  +    {_{f}K}^N_N(\Lambda_N)\\
\end{split}
\end{equation}
By   (75) and (126) in \cite{Dim09}, we have
\begin{equation}
\begin{split}
\|  {_{f}I}_N(\sigma_N, \Lambda_N)  -1 \|_N   &\leq  4 c^{-1}h^2 |\sigma_N|  \leq    4 c^{-1} h^2   r2^{-N} \\
\|{_{f}K}^N_N(\Lambda_N)\|_N   &\leq A^{-1}  \|{_{f}K}^N_N \|_N  \leq  A^{-1}r 2^{-N}\\
\end{split}
\end{equation} 
so that  $\| {_{f}F}_N (\Lambda_N) \|_N \leq \left(  4c^{-1} h^2 + A^{-1} \right)r (2^{-N})$ which  is   ${{\cal O}}(2^{-N})$  as  $N \to \infty$.

In  lemma 14 \cite{Dim09}) Dimock has proved that  for  $h$  sufficiently large 
\begin{equation}
\int   G_N(\Lambda_N,0, \zeta)  d \mu_{C_N} (\zeta)  \leq     2 
\end{equation}
Then we estimate
\begin{equation}
\begin{split} \label{tailbound}
\left|\int        {_{f}F}_N(\Lambda_N ) d \mu_{ C_N}\right | & \leq        \|  {_{f}F}_N(\Lambda_N) \|_N \int   G_N(\Lambda_N,0, \zeta)
  d \mu_{C_N} (\zeta)  \\
&\leq  2  \|  F(\Lambda_N) \|_N  \\
& \leq 2\left(  4c^{-1} h^2 + A^{-1} \right)r (2^{-N}) \\
\end{split}
\end{equation}
Hence   the expression  (\ref{153})  is  $  {{\cal O}}(2^{-N})   | \Lambda_N|^{-1} $
 and  goes to zero very  quickly  as  $N  \to  \infty$ 
 
The rest of the proof came  as in the proof of Theorem 8 in \cite{Dim09}.

\section{Correlation functions: estimates and infinite volume limit}    \label{result1}

Note: We always can assume that $L \gg 2^{d+3} + 1$

\subsection{In the case: $f(\phi) = \sum_{k=1}^m t_k \partial_{\mu_k} \phi (x_k)$}

For   $x_k \in {\mathbb{Z}}^d$ are different points; $\mu_k \in \{\pm1,\dots,\pm d \}$ and $t_k$  complex and $|t_k| \leq a = a(r)$ for $\forall k: 1,2,... m$.
\subsubsection{Proof of  Theorem \ref{corrpa}}
Using (\ref{afterflow}) with $l = N$, for the truncated correlation functions, we have:
\begin{equation} \label{truncated}
\begin{split}
\cG^t &(x_1, x_2,\dots x_m) \equiv  \big<\prod_{k=1}^m\partial_{\mu_k} \phi(x_k) \big>^t  \equiv i^m \frac{\partial^m}{\partial t_1 \dots  \partial t_m} \log {_{f}Z'} \Big|_{t_1=0,\dots t_m=0} \\
= & i^m \frac{\partial^m}{\partial t_1 \dots \partial t_m} \left(  \sum_{j=1}^N \sum_{B \in \cB_j(\Lambda_N)}     {_{f}E}^N_j( B)   \right)   \Big|_{t_1=0,\dots t_m=0} \\
& + i^m \frac{\partial^m}{\partial t_1 \dots \partial t_m} \log  \int   ({_{f}I}_N(\sigma_N) \circ  {_{f}K}^N_N )(\Lambda_N) d   \mu_{C_N} \Big|_{t_1=0,\dots t_m=0} \\
= &  i^m \sum_{j=1}^N \sum_{B \in \cB_j(\Lambda_N)}   \frac{\partial^m}{\partial t_1\dots \partial t_m}   {_{f}E}^N_j( B)      \Big|_{t_1=0,\dots t_m=0} \\
& + i^m\frac{\partial^m}{\partial t_1 \dots  \partial t_m} \log \left( 1 +  \int   ({_{f}I}_N(\sigma_N) -1 +  {_{f}K}^N_N )(\Lambda_N) d   \mu_{C_N}\right)  \Big|_{t_1=0,\dots t_m=0} \\
\end{split}
\end{equation}
Now we consider the quantity:
\begin{equation}
\begin{split}
{_{f}F_N} &\equiv \sum_{j=1}^{N} \sum_{B\in \cB_j(\Lambda_N)} \frac{\partial^m}{\partial{t_1}... \partial{t_m}}  {_{f}E}_j^N (B) \Big|_{t_1=0,... t_m=0} \\
&= \sum_{j=0}^{N-1} \sum_{B\in \cB_j(\Lambda_N)} \frac{\partial^m}{\partial{t_1}... \partial{t_m}} {_{f}\beta}({_{f}K}_j^N,B) \Big|_{t_1=0,... t_m=0} \\
&= \sum_{j=0}^{N-1} \sum_{B\in \cB_j(\Lambda_N)} \frac{\partial^m}{\partial{t_1}... \partial{t_m}} \sum_{X \in \cS_j, X\supset B} \frac{1}{|X|_j} {{_{f}K}_j^N}^{\#}(X,0)  \Big|_{t_1=0,... t_m=0}\\
\end{split}
\end{equation}
by the definition of ${_{f}\beta}$ in (\ref{DefAlBe}).

 We notice that $\frac{\partial^m}{\partial{t_1}... \partial{t_m}}  {_{f}E}_j (B) =0$ unless $B^*  \supset \{ x_1, x_2,... x_m\}$. Therefore,

\begin{equation}
{_{f}F}_N= \sum_{j=0}^{N-1} \sum_{\substack{B\in \cB_j(\Lambda_N)\\B^*  \supset \{ x_1, x_2,... x_m\}}} \frac{\partial^m}{\partial{t_1}... \partial{t_m}} \sum_{X \in \cS_j, X\supset B} \frac{1}{|X|_j} {{_{f}K}_j^N}^{\#}(X,0)  \Big|_{t_1=0,... t_m=0}
\end{equation}

{\bf Note:}  Let $\eta = \min \{d/2, 2\}. $ For any small $\epsilon > 0$,  we can always find $A, L$ sufficiently large such that: 
\begin{equation}
\begin{split}
\| \left( \cL_1 + \cL_2 + \cL_3' + \cL_4     \right) ({_{0}K}) \|_{j+1} &\leq \frac{1}{4L^{\eta -\epsilon}} \| {_{0}K} \|_j \\
\| \left( \cL_1 + \cL_2 \right) (\de {_{f}K}) \|_{j+1} \leq \frac{1}{4L^{\eta-\epsilon}} \left( \|\de {_{f}K} \|_j \right)
&\leq \frac{1}{4L^{\eta-\epsilon}} \left( \| {_{0}K} \|_j + \| {_{f}K} \|_j \right)\\
\end{split}
\end{equation}
with $j\geq 1$ by using the explicit upper bounds in Lemmas \ref{L1}, \ref{L2}, \ref{L3prime}, and \ref{L4}.

Then we can replace $\mu = 1/2$ in Theorem 7 by $\mu =1/ M$ for  $M = L^{\eta-\epsilon} \geq 2$. We still have $|\sigma_j|  \leq  rM^{-j}$  and  $ \| {_{f}K}^N_j\|_j  \leq  rM^{-j}$  and    $\|{_{f}E}^N_{j+1}\|_{j+1}  \leq  {{\cal O}}(L^d)  r M^{-j}$ with $\max_k |t_k| < a$ sufficiently small and $0 \leq j \leq N-1$.
Because ${{_{f}K}_j^N}^{\#}(X,0)$ is analytic, using Cauchy's bound and (79) in \cite{Dim09}, we have:
\begin{equation}
\begin{split}
\Big| \frac{\partial^m}{\partial{t_1}... \partial{t_m}}{ {_{f}K}_j^N}^{\#}(X,0) \big|_{t_1=0,... t_m=0} \Big|
&\leq \frac{m!}{a^m} \left( \frac{A}{2} \right)^{-|X|_j}   \| {_{f}K}^N_j\|_j    \\
&\leq \frac{m!}{a^m} \left( \frac{A}{2} \right)^{-|X|_j} r M^{-j}
\end{split}
\end{equation}
Then
\begin{equation}
\begin{split}
&\Big|  \frac{\partial^m}{\partial{t_1}... \partial_{t_m}} \sum_{X \in \cS_j, X\supset B} \frac{1}{|X|_j} {{_{f}K}_j^N}^{\#}(X,0)  \Big|_{t_1=0,... t_m=0}  \Big|\\
&\leq  \sum_{X \in \cS_j, X\supset B} \frac{1}{|X|_j} \Big|  \frac{\partial^m}{\partial{t_1}... \partial{t_m}} {{_{f}K}_j^N}^{\#}(X,0)  \Big|_{t_1=0,... t_m=0} \Big|  \\
&\leq  \sum_{X \in \cS_j, X\supset B} \frac{1}{|X|_j}   \frac{m!}{a^m} \left( \frac{A}{2} \right)^{-|X|_j} r M^{-j} \\
&\leq n_3(d,\frac{A}{2}) \frac{m!rM^{-j}}{a^m}\\
\end{split}
\end{equation}
So
\begin{equation}
\begin{split}
|{_{f}F_N}| &\leq \sum_{j=0}^{N-1} \sum_{\substack{B\in \cB_j(\Lambda_N)\\B^*  \supset \{ x_1, x_2,... x_m\}}}
 n_3(d,\frac{A}{2}) \frac{m!rM^{-j}}{a^m}\\
\end{split}
\end{equation}
By  (\ref{153})-(\ref{tailbound}), we have:
\begin{equation}\label{corr2tail}
\begin{split}
&\Big| \log \left( 1 +  \int   ({_{f}I}_N(\sigma_N) -1 +  {_{f}K}^N_N )(\Lambda_N) d   \mu_{C_N}\right)  \Big| \\
& \leq \log \left( 1 + 2 \| F(\Lambda_N) \|_N \right) \\
& \leq \log  \left( 1 + 2 [4c^{-1}h^2 + A^{-1}]  r 2^{-N} \right) \\
\end{split}
\end{equation}
Using the Cauchy's bound as above, 
we obtain:
\begin{equation}\label{corr2tail}
\begin{split}
&\Big|  \frac{\partial^m}{\partial t_1... \partial t_m} \log \left( 1 +  \int   ({_{f}I}_N(\sigma_N) -1 +  {_{f}K}^N_N )(\Lambda_N) d   \mu_{C_N}\right)  \Big|_{t_1=0,...t_m=0} \Big|  \\ 
& \leq \frac{m!}{a^m} \log  \left( 1 + 2 [4c^{-1}h^2 + A^{-1}]  r 2^{-N} \right) \\
\end{split}
\end{equation}
So
\begin{equation}\label{corr2tail}
\begin{split}
 \lim_{N \to \infty}&\Big|  \frac{\partial^m}{\partial t_1... \partial t_m} \log \left( 1 +  \int   ({_{f}I}_N(\sigma_N) -1 +  {_{f}K}^N_N )(\Lambda_N) d   \mu_{C_N}\right)  \Big|_{t_1=0,...t_m=0} \Big| = 0\\
\end{split}
\end{equation}

Now let $j_0$ be the smallest integer such that $ \exists B \in \cB_{j_0} : B^* \supset \{ x_1, x_2,... x_m \}$.

Without loosing the generality, we can assume that $|x_1 - x_2| =  \textrm{ diam} (x_1,... x_m) $

For every $j \geq j_0$, let $B^1_j \in \cB_j$ be the unique  $j$-block that contains  $\{ x_1\}$. For any $B \in \cB_j, j \geq j_0$ with $B^* \supset \{ x_1, x_2,... x_m \}$, $B$ must be in $ {B^1_j}^*$. 

We have 
\begin{equation} \label{fFN}
\begin{split}
|{_{f}F_N}| &\leq \sum_{j=0}^{N-1} \sum_{\substack{B\in \cB_j(\Lambda_N)\\ B^*\supset \{ x_1, x_2,... x_m \}}}
 n_3(d,\frac{A}{2}) \frac{m!rM^{-j}}{a^m}\\
&=  \sum_{j=j_0}^{N-1} \sum_{\substack{B\in \cB_j(\Lambda_N)\\ B^*\supset \{ x_1, x_2,... x_m \}}}       n_3(d,\frac{A}{2}) \frac{m!rM^{-j}}{a^m}\\
\end{split}
\end{equation}
Since $M \geq 2$, the last part of (\ref{fFN})  is bounded by
\begin{equation} \label{2ndterm}
\begin{split}
\sum_{j=j_0}^{N-1} \sum_{\substack{B\in \cB_j(\Lambda_N)\\ B^*\supset \{ x_1, x_2,... x_m \}}}
 n_3(d,\frac{A}{2}) \frac{m!rM^{-j}}{a^m}  &\leq \sum_{j=j_0}^{N-1} \sum_{\substack{B\in \cB_j(\Lambda_N)\\B \in {B^1_j}^*}}
 n_3(d,\frac{A}{2}) \frac{m!rM^{-j}}{a^m} \\
&\leq \sum_{j=j_0}^{N-1}(2^{d}2)^d n_3(d,\frac{A}{2}) \frac{m!rM^{-j}}{a^m}\\
&\leq  2^{d(d+1)} n_3(d,\frac{A}{2})2 \frac{m!rM^{-j_0}}{a^m} \\
\end{split}
\end{equation}

Therefore, we have:
\begin{equation}
|{_{f}F_N}| \leq   2^{d(d+1)}  n_3(d,\frac{A}{2})2 \frac{m!rM^{-j_0}}{a^m}\\
\end{equation}

By the definition of $j_0$, we have: $ |x_1 - x_2| \leq d2^{d+1} L^{j_0}$. Because $M = L^{\eta-\epsilon}$, we get 
\begin{equation}
\begin{split}
M^{-j_0} =  L^{-j_0(\eta-\epsilon)} &\leq  (d2^{d+1})^\eta |x_1 - x_2|^{-\eta+ \epsilon}   \\
&=  (d2^{d+1})^\eta  \textrm{ diam}^{-\eta+\epsilon} (x_1,\dots,x_m) \\
\end{split}
\end{equation}
Hence, we have:
\begin{equation}
|{_{f}F_N}| \leq     2^{d(d+1)}  n_3(d,\frac{A}{2})2 \frac{m!r}{a^m}  \textrm{ diam}^{-\eta+\epsilon} (x_1,... x_m) \left(  d^\eta 2^{\eta(d+1)} \right)  
\end{equation}
Using this with (\ref{corr2tail}), we obtain:
\begin{equation}
\begin{split}
&\Big| \frac{\partial^m}{\partial t_1 ...  \partial t_m} \log {_{f}Z'} \Big|_{t_1=0,...t_m=0} \Big| 
\leq    2^{d(d+1)} 4 n_3(d,\frac{A}{2}) \frac{m!r}{a^m}  \textrm{ diam}^{-\eta+\epsilon} (x_1,... x_m) \left(  d^\eta 2^{\eta(d+1)} \right)   \\
\end{split}
\end{equation}
Combining with (\ref{n3dl}), we get  $n_3(d,\frac{A}{2} )2^{d(d+1)} 4r  (d2^{d+1})^\eta\leq 1$ with sufficiently large $A$. Therefore, with sufficiently large $A$,  we have: 
\begin{equation}
\begin{split}
\Big|  \cG^t (x_1, x_2,... x_m) \Big| &= \Big|  \big<\prod_{k=1}^m\partial_{\mu_k} \phi(x_k) \big>^t \Big|  =
\Big| \frac{\partial^m}{\partial t_1 ...  \partial t_m} \log {_{f}Z'} \Big|_{t_1=0,...t_m=0} \Big| \\
&\leq \frac{m!}{a^m}  \textrm{ diam}^{-\eta+\epsilon} (x_1,... x_m) \\
\end{split}
\end{equation}
We complete the proof of Theorem 2.

%\bigskip
{\noindent{\bf Remark}. }Actually for any $N-1 \geq q\geq j_0$, similarly to (\ref{2ndterm}), we have
\begin{equation} \label{boundtailseries}
\begin{split}
&\Big | \sum_{j=q}^{N-1} \sum_{\substack{B\in \cB_j(\Lambda_N)\\B^*  \supset \{ x_1, x_2,... x_m\}}} \frac{\partial^m}{\partial{t_1}... \partial{t_m}} \sum_{X \in \cS_j, X\supset B} \frac{1}{|X|_j} {{_{f}K}_j^N}^{\#}(X,0)  \Big|_{t_1=0,... t_m=0} \Big|   \\
&\leq \sum_{j=q}^{N-1} \sum_{\substack{B\in \cB_j(\Lambda_N)\\ B^*\supset \{ x_1, x_2,... x_m \}}}
 n_3(d,\frac{A}{2}) \frac{m!rM^{-j}}{a^m}\\
&\leq  2^{d(d+1)} n_3(d,\frac{A}{2})2 \frac{m!rM^{-q}}{a^m} \\
\end{split}
\end{equation}

\subsubsection{Proof of  Theorem \ref{inflimcorr}}
Now we fix the set $\{x_1, x_2,... x_m \}$. Let $j_1$ be the smallest integer such that $B_{j_1}^0 \supset \{ x_1, x_2,... x_m \}$.  Then $j_1$ is the smallest integer which is greater than $\log_L \max_i \| x_i \|_\infty$. We also have:  $j_0 \leq j_1$.

Let $q$ be any number such that $q \geq  j_1 +1 \geq j_0 +1$. And let $N_1, N_2$ be any integers such that $N_2 \geq N_1 > q $. Using the definition of $j_0$, we have
\begin{equation}
\begin{split}
{_{f}F}_{N_1}=&\sum_{j=j_0}^{q-1} \sum_{\substack{B\in \cB_j(\Lambda_N)\\B^*  \supset \{ x_1, x_2,... x_m\}}} \frac{\partial^m}{\partial{t_1}... \partial{t_m}} \sum_{X \in \cS_j, X\supset B} \frac{1}{|X|_j} {{_{f}K}_j^{N_2}}^{\#}(X,0)  \Big|_{t_1=0,... t_m=0}\\
+&\sum_{j=q}^{N_1-1} \sum_{\substack{B\in \cB_j(\Lambda_N)\\B^*  \supset \{ x_1, x_2,... x_m\}}} \frac{\partial^m}{\partial{t_1}... \partial{t_m}} \sum_{X \in \cS_j, X\supset B} \frac{1}{|X|_j} {{_{f}K}_j^{N_2}}^{\#}(X,0)  \Big|_{t_1=0,... t_m=0}\\
%\end{split}
%\end{equation}
%and
%\begin{equation}
%\begin{split}
{_{f}F}_{N_2}=&\sum_{j=j_0}^{q-1} \sum_{\substack{B\in \cB_j(\Lambda_N)\\B^*  \supset \{ x_1, x_2,... x_m\}}} \frac{\partial^m}{\partial{t_1}... \partial{t_m}} \sum_{X \in \cS_j, X\supset B} \frac{1}{|X|_j} {{_{f}K}_j^{N_2}}^{\#}(X,0)  \Big|_{t_1=0,... t_m=0}\\
+&\sum_{j=q}^{N_2-1} \sum_{\substack{B\in \cB_j(\Lambda_N)\\B^*  \supset \{ x_1, x_2,... x_m\}}} \frac{\partial^m}{\partial{t_1}... \partial{t_m}} \sum_{X \in \cS_j, X\supset B} \frac{1}{|X|_j} {{_{f}K}_j^{N_2}}^{\#}(X,0)  \Big|_{t_1=0,... t_m=0}\\
\end{split}
\end{equation}
We also notice that:
\begin{equation}
\begin{split}
&\sum_{j=j_0}^{q-1} \sum_{\substack{B\in \cB_j(\Lambda_N)\\B^*  \supset \{ x_1, x_2,... x_m\}}} \frac{\partial^m}{\partial{t_1}... \partial{t_m}} \sum_{X \in \cS_j, X\supset B} \frac{1}{|X|_j} {{_{f}K}_j^{N_2}}^{\#}(X,0)  \Big|_{t_1=0,... t_m=0}\\
&= \sum_{j=j_0}^{q-1} \sum_{\substack{B\in \cB_j(\Lambda_N)\\B^*  \supset \{ x_1, x_2,... x_m\}}} \frac{\partial^m}{\partial{t_1}... \partial{t_m}} \sum_{X \in \cS_j, X\supset B} \frac{1}{|X|_j} {{_{f}K}_j^{N_1}}^{\#}(X,0)  \Big|_{t_1=0,... t_m=0}\\
\end{split}
\end{equation}
because for $0 \leq j \leq q-1$, ${{_{f}K}_j^N}^{\#}(X,0) $ only depend on $\phi$ within $X^*$ and $X^* \subset \Lambda_q $ which is  the center $q$-block of $\Lambda_{N_1} \subset \Lambda_{N_2}$. 
Therefore,
\begin{equation}
\begin{split}
|{_{f}F}_{N_2} - {_{f}F}_{N_1}| 
&\leq \Big| \sum_{j=q}^{N_2-1} \sum_{\substack{B\in \cB_j(\Lambda_N)\\B^*  \supset \{ x_1, x_2,... x_m\}}} \frac{\partial^m}{\partial{t_1}... \partial{t_m}} \sum_{X \in \cS_j, X\supset B} \frac{1}{|X|_j} {{_{f}K}_j^{N_2}}^{\#}(X,0)  \Big|_{t_1=0,... t_m=0} \Big|  \\
&+ \Big| \sum_{j=q}^{N_1-1} \sum_{\substack{B\in \cB_j(\Lambda_N)\\B^*  \supset \{ x_1, x_2,... x_m\}}} \frac{\partial^m}{\partial{t_1}... \partial{t_m}} \sum_{X \in \cS_j, X\supset B} \frac{1}{|X|_j} {{_{f}K}_j^{N_1}}^{\#}(X,0)  \Big|_{t_1=0,... t_m=0} \Big|  \\
\end{split}
\end{equation}

Then using (\ref{boundtailseries}) with $\mu = 1/2$ instead of $\mu =1/M = L^{-\eta + \epsilon}$, we obtain:
\begin{equation}
\begin{split}
\Big| \sum_{j=q}^{N_2-1} \sum_{\substack{B\in \cB_j(\Lambda_N)\\B^*  \supset \{ x_1, x_2,... x_m\}}}& \frac{\partial^m}{\partial{t_1}... \partial{t_m}} \sum_{X \in \cS_j, X\supset B} \frac{1}{|X|_j} {{_{f}K}_j^{N_2}}^{\#}(X,0)  \Big|_{t_1=0,... t_m=0} \Big| \\
&\leq  2^{d(d+1)} n_3(d,\frac{A}{2}) 2 \frac{m!r2^{-q}}{a^m}\\
\end{split}
\end{equation}
and
\begin{equation}
\begin{split}
 \Big| \sum_{j=q}^{N_1-1} \sum_{\substack{B\in \cB_j(\Lambda_N)\\B^*  \supset \{ x_1, x_2,... x_m\}}}& \frac{\partial^m}{\partial{t_1}... \partial{t_m}} \sum_{X \in \cS_j, X\supset B} \frac{1}{|X|_j} {{_{f}K}_j^{N_1}}^{\#}(X,0)  \Big|_{t_1=0,... t_m=0} \Big|\\
&\leq  2^{d(d+1)} n_3(d,\frac{A}{2}) 2 \frac{m!r2^{-q}}{a^m}\\
\end{split}
\end{equation}
That means we have:
\begin{equation}
|{_{f}F}_{N_2} - {_{f}F}_{N_1}| \leq  2^{d(d+1)} n_3(d,\frac{A}{2}) 4 \frac{m!r2^{-q}}{a^m} \to 0
\end{equation}
when ${q\to \infty}$. \\
Combining this with (\ref{truncated}) and (\ref{corr2tail}), we can conclude that $\lim_{N \to \infty}  \big<\prod_{k=1}^m \partial_{\mu_k} \phi(x_k)\big>^t $  exists.

 \bigskip

{\noindent{\bf Remark}. }We have $N$-uniformly boundedness on correlation functions and 

$\lim_{N \to \infty} \cG^t  (x_1, x_2,... x_m)$ exists. Therefore the bounds  are held for infinite volume limit

\subsection{When  $f(\phi) = \sum_{k=1}^m t_k \exp \left( i\partial_{\mu_k} \phi(x_k) \right)$}

Using exactly the same argument as the above subsection, we obtain these following results:

\begin{thm} \label{CorrExp}
For any small $\epsilon > 0$, with $ L, A$ sufficiently large (depending on $\epsilon$), let $\eta = \min \{ d/2, 2 \}$ we have:    
\begin{equation}
\begin{split}
\Big| \big<\prod_{k=1}^m \exp \left( i\partial_{\mu_k} \phi(x_k) \right) \big>^t \Big|  \leq \frac{m!}{a^m}  \textrm{ diam}^{-\eta+\epsilon} (x_1,... x_2)  \\ 
\end{split}
\end{equation}
where $a$ depends on $\epsilon, L, A$
\end{thm}
% \bigskip

\begin{thm} \label{inflimCorrExp} %For any fixed set $\{x_1, x_2,.. x_m\} \subset {\mathbb{Z}}^d$, 
With $L, A$ sufficiently large, the infinite volume limit of the truncated correlation function $\lim_{N \to \infty}  \big<\prod_{k=1}^m \exp \left( i\partial_{\mu_k} \phi(x_k) \right) \big>^t $ exists
\end{thm}

\subsection{Other cases} 

We can consider $f (\phi) = \sum_{k=1}^m t_k f_k( \phi) ( x_k)$ with
 
* $t_k \in {\mathbb{C}}$

* $x_k \in {\mathbb{Z}}^d$ are different points.

* $f_k$ is bounded in the sense that there are some $M_k, m_k \geq 0$ such that 
\begin{equation}
\| f_k ( \{ x_k \}, \phi) \|_0 \leq M_k \|\phi\|_{\Phi_0} + m_k
\end{equation}

With the same argument as above cases, we have:
\begin{thm} \label{Corrgen}
For any small $\epsilon > 0$, with $ L, A$ sufficiently large (depending on $\epsilon$), let $\eta = \min \{ d/2, 2 \}$ we have:    
\begin{equation}
\begin{split}
\Big| \big< \prod_{k=1}^m  f_k( \phi) ( x_k) \big>^t \Big|  \leq \frac{m!}{a^m}  \textrm{ diam}^{-\eta+\epsilon} (x_1,... x_2)  \\ 
\end{split}
\end{equation}
where $a$ depends on $\epsilon, L, A$
\end{thm}

\begin{thm} \label{inflimCorrgen}
With $L, A$ sufficiently large, the infinite volume limit of the truncated correlation function $\lim_{N \to \infty}  \big< \prod_{k=1}^m  f_k( \phi) ( x_k)  \big>^t $ exists
\end{thm}

In the case $f = \sum_{k=1}^m t_k W_0 (\{x_k\})$, with $W_0(\{x_k\})  =  z W(1,  \{x_k \})$  as  in    (\ref{defWuB}). Using the Lemma \ref{WuB} (or the lemma 4 in \cite{Dim09} ), these $W_0 (\{x_k\})$ satisfy those above conditions. 
 The  $W_0(\{x_k\})$ are  actually the density of the dipoles at $x_k$ used in \cite{BryKel93}. Applying theorems \ref{Corrgen} and \ref{inflimCorrgen}, we obtain these results:

\begin{cor} \label{CorrDens}
For any small $\epsilon > 0$, with $ L, A$ sufficiently large (depending on $\epsilon$), let $\eta = \min \{ d/2, 2 \}$ we have:    
\begin{equation}
\begin{split}
\Big| \big< \prod_{k=1}^m W_0 ( \{ x_k\}) \big>^t \Big| \leq \frac{m!}{a^m}  \textrm{ diam}^{-\eta+\epsilon} (x_1,... x_2)  \\ 
\end{split}
\end{equation}
\end{cor}
This result somehow looks like the theorem (1.1.2) in \cite{BryKel93}. However it gives estimates for truncated correlation functions of $(p \geq 2)$ points instead of some  estimate for only 2 points.

\begin{cor}  \label{inflimCorrDens}
With $L, A$ sufficiently large, the infinite volume limit of the truncated correlation function $\lim_{N \to \infty}  \big< \prod_{k=1}^m  W_0 ( \{ x_k\})  \big>^t $  exists
\end{cor}

{\noindent{\bf Remark}. }We can consider the more general form  $f (\phi) = \sum_{k=1}^m t_k f_k (\phi)$ with
 
* $t_k \in {\mathbb{C}}$

* $A_k \equiv { \mathrm{ supp  }} f_k$ are pairwise disjoint and $|A_k| < \infty$

* $f_k$ is bounded in the sense that there are some $M_k, m_k \geq 0$ such that 
\begin{equation}
\| f_k (A_k,\phi)\|_0 \leq M_k \|\phi\|_{\Phi_0} + m_k
\end{equation}

Then we still get similar results as in Theorems \ref{Corrgen} and \ref{inflimCorrgen}.

\appendix  
\section{Kac-Siegert Transformation}  \label{KacSie}

 By expanding the exponential  in  (\ref{SineGordon})   and carrying  out the Gaussian integrals, we can rewrite ${_{0}Z}_N$ as
\begin{equation}
\begin{split}
{_{0}Z}_N &=\int \left(\sum_{n\geq 0} \frac{z^n}{n!}\prod_{i=1}^{n} \sum_{x_i \in  \Lambda_N \cap  {\mathbb{Z}}^d}  \int_{{\mathbb{S}}^{d-1}}  dp_i  (e^{ip_i \cdot \partial \phi (x_i) } + e^{-ip_i \cdot \partial \phi (x_i)})/2 \right)  d \mu_{C} ( \phi  )\\
&=\int \left(\sum_{n\geq 0} \frac{z^n}{n!}\prod_{i=1}^{n} \sum_{x_i \in  \Lambda_N \cap  {\mathbb{Z}}^d}  \int_{{\mathbb{S}}^{d-1}}  dp_i e^{ip_i \cdot \partial \phi (x_i) } \right)  d \mu_{C} ( \phi  )\\
&=  \sum_{n\geq 0} \frac{z^n}{n!} \prod_{i=1}^n  \sum_{x_i \in  \Lambda_N \cap  {\mathbb{Z}}^d}  \int_{ {\mathbb{S}}^{d-1}}dp_i 
\int e^{i\sum_{k=1}^n p_k \cdot \partial \phi (x_k) }d \mu_{C} ( \phi  )\\
&=  \sum_{n\geq 0} \frac{z^n}{n!} \prod_{i=1}^n  \sum_{x_i \in  \Lambda_N \cap  {\mathbb{Z}}^d}  \int_{{\mathbb{S}}^{d-1} }   dp_i \exp \left( \frac{-1}{2} \sum_{1 \leq k,j \leq n} (p_k \cdot \partial)(p_j \cdot \partial) C (x_k, x_j)   \right)
\end{split}
\end{equation}
which is exactly the same as the  grand canonical  partition  function (\ref{grand}).

\bigskip

{\bf{\large Acknowledgments} }

This work was in partial fulfillment of the requirements for the Ph.D. degree at the University at Buffalo, State University of New York. The author owes deep gratitude to his Ph.D. advisor Jonathan D. Dimock for his continuing help and support. Dimock's prior investigations of {\em infinite volume limit for the dipole gas} \cite{Dim09}  have served  as a framework for much of the current work.

\end{document}